\documentclass[letterpaper,twocolumn,10pt]{article}

\usepackage{graphicx}
\usepackage{subfig}
\usepackage{times}
\usepackage{microtype}
\usepackage{textcomp}
\usepackage{multirow}
\usepackage{framed}
\usepackage{amsmath}
\usepackage{listings}
\usepackage{flushend}
\usepackage{authblk}
\PassOptionsToPackage{hyphens}{url}
\usepackage{hyperref}
\usepackage{enumitem}
\usepackage{amssymb}

\usepackage{comment}
\usepackage{pgfplots}
\usepackage{xspace}
\newcommand{\sysname}{Il\'{u}vatar\xspace}

\usepackage{etoolbox}
\gappto{\UrlBreaks}{\UrlOrds}

\hypersetup{breaklinks=true}

\title{FaasMeter: Energy-First Serverless Computing}

\author[1]{Abdul Rehman}

\author[1]{Alexander Fuerst}

\author[1]{Prateek Sharma}
\affil[1]{Indiana University}

\begin{document}
\date{}

\maketitle 

\begin{abstract}

Functions as a Service has emerged as a popular abstraction for a wide range of cloud applications and an important cloud workload. We present the design and implementation of FaasMeter, a FaaS control plane which provides energy monitoring, accounting, control, and pricing as first-class operations. The highly diverse and dynamic workloads of FaaS create additional complexity to measuring and controlling energy usage which FaasMeter can mitigate. 

We develop a new statistical energy disaggregation approach to provide accurate and complete energy footprints for functions, despite using noisy and coarse-grained system-level power (not just CPU power readings). Our accurate and robust footprints are achieved by combining conventional power models with Kalman filters and Shapley values. FaasMeter is a full-spectrum energy profiler, and fairly attributes energy of shared resources to  functions (such as energy used by the control plane itself). We develop new energy profiling validation metrics, and show that FaasMeter's energy footprints are accurate to within 1\% of carefully obtained marginal energy ground truth measurements.

\end{abstract}

\vspace*{-0.3cm}
\section{Introduction}
\vspace*{-0.3cm}

Energy is a key resource in large scale distributed computing. 
Cloud platforms consume a significant ($>1\%$) amount of global energy~\cite{Siddik_2021}, and reducing it is a vital step in IT decarbonization efforts~\cite{mghpcc-ic, agarwal_redesigning_2021, acun_carbon_2022, anderson_treehouse_2022}.
While energy informs the design, operation, costs, and capabilities of cloud infrastructure and data centers, cloud \emph{applications} are still mostly oblivious of energy as the primary underlying resource.
With a growing awareness and need for environmental sustainability, the energy and carbon footprint of cloud applications will increasingly serve as the primary accounting and optimization metric~\cite{souza2023ecovisor}, complementing the traditional metrics such as monetary cost and resource utilization (e.g., CPU). 
This requires energy and carbon awareness throughout the cloud computing stack---hardware, resource allocation software, applications, etc.
However, cloud abstractions and applications are continuously evolving---making energy observability and optimization a moving target.

Serverless computing, or Functions as a Service (FaaS), has emerged as a key cloud abstraction which is enabling rapid application development and  deployment~\cite{serverless-cacm-21, castro2019rise, adzic2017serverless}.
Cloud functions are small, self-contained programs, whose entire execution is managed by the cloud provider. 
They have low cost, auto-scaling, and a ``serverless'' model where users don't have to worry about explicit resource management. 
Serverless computing is a major and growing cloud workload~\cite{shahrad_serverless_2020}, and serves as the resource abstraction for a wide range of event-driven applications (such as web and API services, IoT, and ML inference), workflows~\cite{funcx_hpdc_20, mahgoub_wisefuse_2022}, and even  throughput-intensive parallel workloads~\cite{xu2021lambda,carreira_cirrus_2019,fouladi_laptop_2019, fouladi2017encoding}.
FaaS is also an increasingly useful abstraction for harnessing computational accelerators~\cite{du_serverless_2022, maschi_serverless_2023, yu_faaswap_2023} and edge computing resources~\cite{wang2021lass}.

While the \emph{performance} of serverless functions has received significant attention, in this paper we take the first step towards quantifying their energy use. 
FaasMeter is our system which provides fine-grained full-system energy observability and control for serverless functions, and virtualizes power in the FaaS control plane.
It elevates energy to a first class resource for serverless computing, and to the best of our knowledge, is the first work on FaaS energy usage. 

The key prerequisite for power virtualization is accurate measurement and attribution. 
However, energy profiling of functions is challenging due to their highly heterogeneous and concurrent nature, and is a major focus of this paper. 
Measuring the energy consumption of applications presents many general challenges which we inherit, along with many FaaS-specific challenges which we identify and address. 
Energy is a shared, global resource, with limited hardware measurement support, especially at the full-system level.
\emph{How do we disaggregate the global energy usage and attribute it to a large number (10s to 100s) of concurrently running heterogeneous functions?}
For complete energy accounting, the shared resources such as the FaaS control plane (e.g., OpenWhisk~\cite{openwhisk}) is also an important component, and its energy consumption must be fairly divided among functions.
Function energy is diffused over space and time, which makes attributing and controlling energy consumption from noisy coarse-grained system-wide power measurements particularly challenging.

Existing power profiling tools, such as Scaphandre~\cite{scaph}, PowerAPI~\cite{fieni_smartwatts_2020}, and others~\cite{he_energat_2023, jay_experimental_2023} are not suitable for FaaS workloads and have many shortcomings.
They rely on CPU power measurements (e.g., RAPL) and do not consider system-wide energy; do not account for shared resources; do not scale well to large number of concurrent functions and CPU cores; and are not sufficiently validated against an external ground truth, rendering them inaccurate and fragile.

We address these shortcomings and challenges by developing an energy metrology framework with new validation methodologies and metrics, and energy benchmarking datasets.
FaasMeter leverages the unique characteristics of FaaS workloads, such as the repeated invocations of functions and their similar resource usage across invocations.
This facilitates statistical disaggregation techniques that use repeated measurements to estimate function per-invocation energy footprints. 
Our approach combines the traditional direct attribution and model-based disaggregation found in existing energy profilers. 
FaasMeter uses various power sources (such as server BMC/IPMI, plug-level meters, and CPU RAPL counters) and OS and control plane metrics, for its statistical inference. 
We perform continuous energy profiling, and develop a Kalman filter approach for adjusting function energy estimates based on workload and system dynamics. 
FaasMeter provides a breakdown of the total energy footprint of functions that includes their fair-share of shared software and hardware components, using the conceptual framework of Shapley values~\cite{winter2002shapley, dong_rethink_2014}.

One of our goals is to enable FaaS providers to be able to charge functions directly for their energy consumption, and thus FaasMeter's energy footprints exhibit the desired fair pricing properties from economics such as proportionality, accuracy, efficiency (i.e., completeness), stability, symmetry, and linearity. 
FaasMeter is integrated into a FaaS control plane, \sysname~\cite{il76-hpdc23}, and provides continuous low-overhead energy profiling and control capabilities such as power capping.
The control of full-system power is increasingly relevant in heterogeneous computing, where CPUs and GPUs are used for running functions and contribute to system-wide energy consumption.
FaasMeter provides \emph{complete} energy footprints, which represent the functions' contribution to the entire server energy. 
The fair energy-pricing for functions enables energy visibility and incentives for both FaaS users and providers.

Validating the energy footprints of individual functions is a major challenge and we develop new methodology and metrics for both internal and external validity.
In particular, we measure the \emph{marginal} energy contribution of a function by running nearly identical workloads, and compare against this ground truth. 
We evaluate FaasMeter on a range of internal and external validation metrics, across different heterogeneous and dynamic workloads, and on different hardware devices.
FaasMeter uses these energy footprints and virtualizes power in the FaaS control plane, and incorporates new capabilities of energy capping and pricing.

\noindent With the goal of making serverless functions energy aware and energy efficient, we make the following contributions:
\begin{enumerate}[wide,labelwidth=!,labelindent=0pt,topsep=0pt,itemsep=-1ex,partopsep=1ex,parsep=1ex]
\item We establish an energy metrology framework for FaaS comprising of new measurement techniques for full-system power, new validation metrics, and  benchmarking datasets. This results in a robust, practical energy management framework for FaaS. 
\item Our energy profiling combines direct and model-based disaggregation to provide \emph{accurate and complete} energy footprints for functions, and meets the Shapley value requirements, making them a viable option for energy pricing of functions.
\item We have extensively evaluated the internal and external validity of FaasMeter on workloads derived from the Azure Function traces~\cite{shahrad_serverless_2020} on different hardware platforms. Our energy footprints are accurate to within 99\% of the marginal energy ground truth. %
\item FaasMeter introduces energy-awareness into the FaaS control plane, and provides energy accounting and control capabilities for system-wide power, and not just CPUs---making it suitable for modern heterogeneous hardware platforms. The function footprint-aware software power-capping achieves an overshoot of less than 3\%.  It is open-source and written in Rust and Python in around 6,000 lines of code. 
\end{enumerate}

\vspace*{-0.3cm}
\section{Background}
\vspace*{-0.3cm}
\label{sec:bg}

FaasMeter builds on and extends two main conceptual components: Functions as a Service and energy-first cloud computing.
This section provides their background and our high-level approach for integrating them.

\vspace*{-0.4cm}
\subsection{Functions as a Service (FaaS)}
FaaS allows users to register small snippets of function code that get executed in response to some event or trigger (such as an HTTP request, message queue event, etc.)~\cite{serverless-cacm-21, aws-lambda, google-functions,azure-functions}. 
Functions are executed inside virtual execution environments such as lightweight hardware virtual machines~\cite{firecracker-nsdi20} or OS containers.
Cloud functions are ``pay for what you use'', and their cost is a combination of their maximum memory allocation and their execution duration~\cite{aws-lambda}. 
One of our goals is to instead introduce \emph{energy-based} pricing where function invocations are charged based on their net energy (or carbon) footprint.

\begin{figure}[t]
  \centering
  \includegraphics[width=0.4\textwidth]{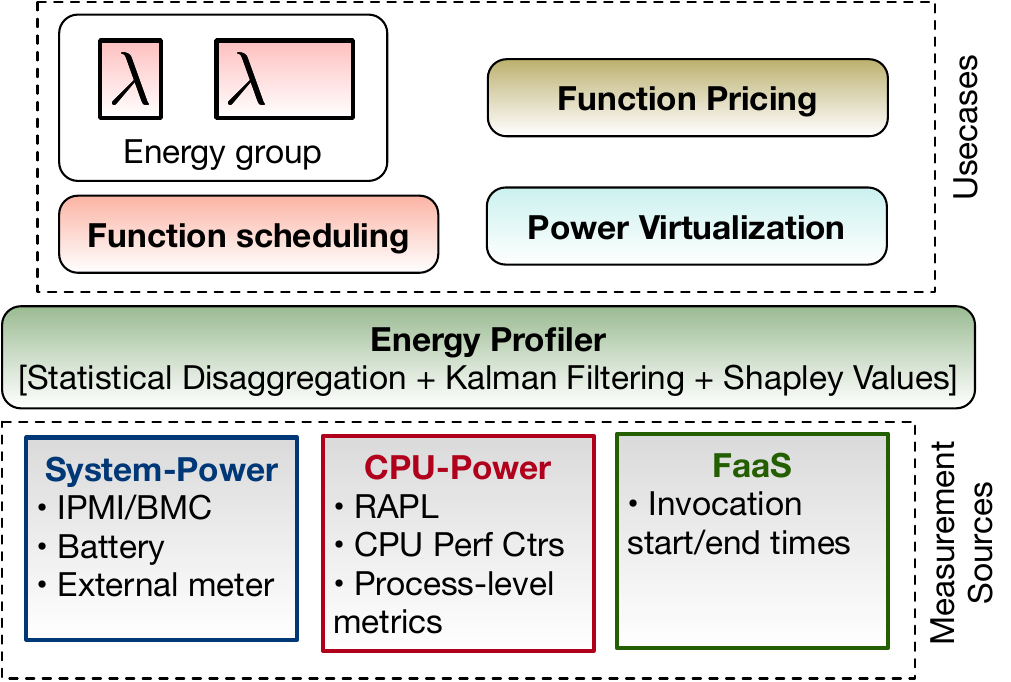}
  \caption{FaasMeter integrates energy-awareness into the FaaS control plane. Multiple power sources are used to obtain accurate per-function energy footprints, which we use for energy capping and pricing.}
  \vspace*{-5pt}
  \label{fig:design}
  \vspace*{-5pt}
\end{figure}

\noindent \textbf{FaaS control planes}  (such as OpenWhisk~\cite{openwhisk})  handle all aspects of function execution. 
They manage the cluster of servers to run functions on, and implement function scheduling, load-balancing, resource monitoring, function status tracking, storing function results, logging, etc. 
They are also responsible for performance optimizations such as keep-alive~\cite{faascache-asplos21} to mitigate function cold-start overheads due to the sandboxing and function initialization overheads. 
Similar to operating systems, they are an important \emph{shared resource} for functions. 
FaaS control planes are highly distributed with many components such as API gateways, distributed message queues (such as Kafka), and databases. %
Current research and production FaaS control planes are energy-oblivious, and do not incorporate any energy management functionality.

\noindent \textbf{Workloads.}
Functions are a common abstraction for accessing cloud resources, and are being used for diverse applications such as web-services, ML inference and training, data analysis, parallel and scientific computing, etc~\cite{serverless-cacm-21, castro2019rise, adzic2017serverless, funcx_hpdc_20, mahgoub_wisefuse_2022}.  %
This results in high workload diversity in all dimensions: the CPU, memory requirements, and inter-arrival-times in public clouds such as Azure are heavy tailed~\cite{shahrad_serverless_2020}.
For example, the inter-arrival times of functions in Azure can range from 0.01 s--1 day, and their execution times can range from 0.1 s to 100 s. 
This also translates to diverse function energy footprints. 
Functions are also popular in edge computing~\cite{cloudflare-workers}, and thus even their execution environments are heterogeneous.
The highly heterogeneous, concurrent, and dynamic nature of FaaS workloads makes it challenging to achieve low latency and high utilization.

\vspace*{-0.4cm}
\subsection{Energy-first Cloud Computing}
\label{sec:bg:energy-aware}

Energy as a first-class resource for operating systems is a long-standing problem and vision~\cite{bellosa2000benefits, flinn_powerscope_1999, zeng_ecosystem_2002, flinn_energy-aware_1999, zeng_currentcy_2003}.
Power virtualization entails accurate process or application level energy measurement~\cite{shen_power_2013}, and fair attribution of shared energy consumers such as the operating system~\cite{ghanei_os-based_2016, ghanei_os-based_2019}. 
Our goal with FaasMeter is to instrumenting a FaaS control-plane and virtualizing power within it, like in~\cite{guo_power_2018}.

\noindent \textbf{Measurement} and observability into energy usage of applications is the first step towards power virtualization, and is increasingly important for environmentally sustainable cloud systems design and implementation. 
Major public clouds are now offering carbon footprint tools for certain cloud applications~\cite{aws_carbon_tool, gcp_carbon_22, azure_sustainability_calc}.
Given this trend, fine-grained energy footprints of serverless applications will be essential for developing energy-aware cloud applications. 

Resource multiplexing is the main challenge for accurate measurement of the power/energy footprint of applications.
Energy is a shared, global resource, and can often only be measured at a coarse granularity both in space and time. 
New hardware capabilities such as RAPL~\cite{zhang_quantitative_2015, khan_rapl_2018}, can provide CPU energy (and in some cases, DRAM~\cite{desrochers_validation_2016}). 
``Software power meters'' such as \texttt{powertop} and others~\cite{do_ptop_2009, colmant_next_2018, fieni_smartwatts_2020, fieni_selfwatts_2021, zhang_estimating_2020, noureddine2013review, khan_energy_2015, wagner_energy_2023, schmitt_online_2019}, use statistical models to attribute total CPU power to processes based on resource use (such as CPU performance counters). 
Modern hardware still only has rudimentary support for power measurement.
Component-level power (such as for network cards) is usually unavailable.
Full-system power can be obtained using server BMCs (baseboard management controllers), battery controllers in mobile devices, external plug-level power meters, or special server hardware~\cite{lee_cloudsocket_2018, ge_powerpack_2010}. 
Power measurement and modeling thus continues to be dominated by CPU-power, but even the accuracy and fidelity of CPU power monitoring remains low, with large jitter and temporal errors~\cite{khan_rapl_2018, anand_hotcarbon22, jay_experimental_2023, ournani2020taming}.

\begin{figure}
    \subfloat[System and CPU power is noisy and coarse-grained. \label{fig:pow-ml-train}]
    {\includegraphics[width=0.22\textwidth]{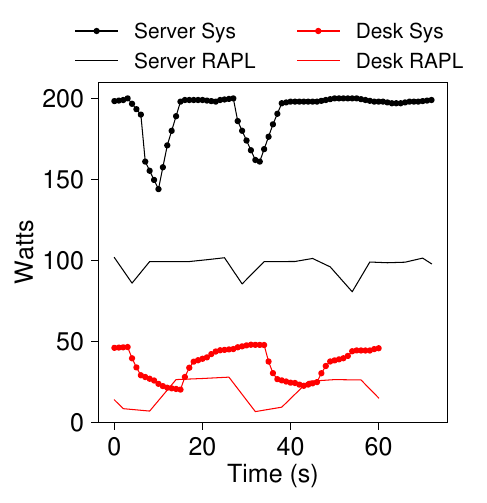}}
    \subfloat[Predicted function footprint with PowerAPI~\cite{noureddine2013review} does not correlate with use (i.e., function invocations).  \label{fig:pow-boundary}]    
{\includegraphics[width=0.24\textwidth]{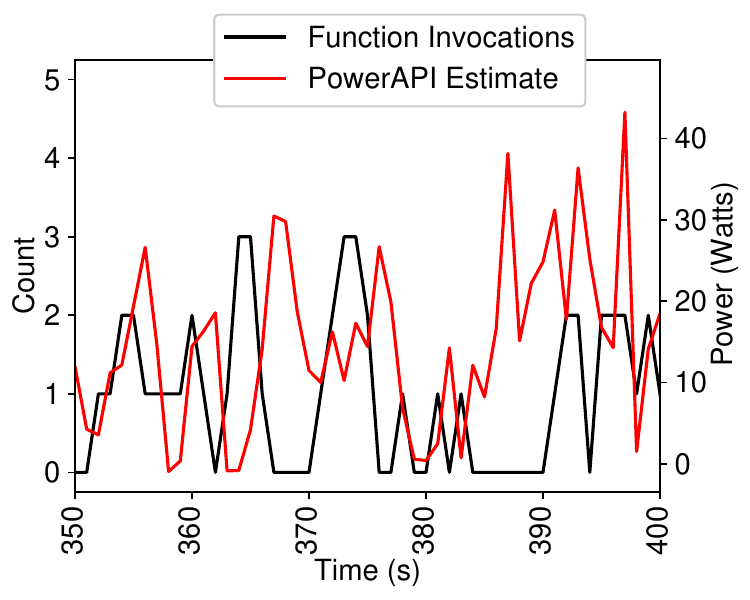}}
     \vspace*{-6pt}
     \caption{Function power signatures cannot be captured reliably by existing power profiling methods.}
     \vspace*{-6pt}
\end{figure}

\begin{figure*}
  \centering
  \includegraphics[width=0.65\textwidth]{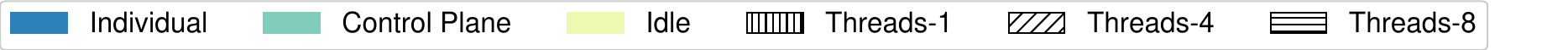}
    \vspace*{-0.7cm}
\end{figure*}
\begin{figure*}
  \centering
  \subfloat[Server-Small \label{subfig:server-small}] 
{\includegraphics[width=0.24\textwidth]{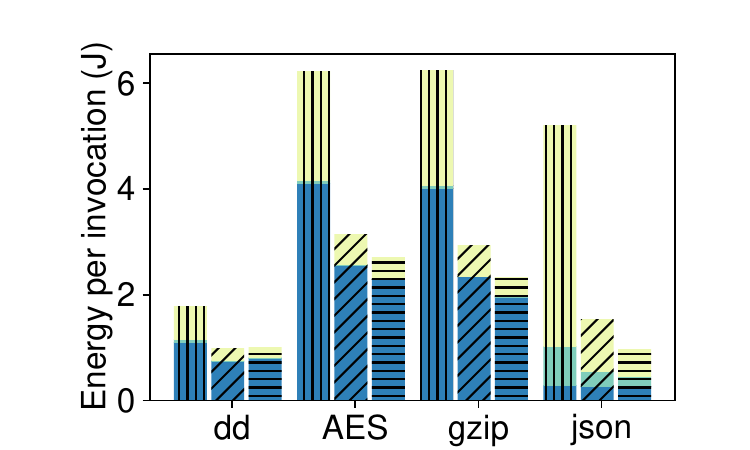}}
  \subfloat[Server-Large \label{subfig:server-large}] 
{\includegraphics[width=0.24\textwidth]{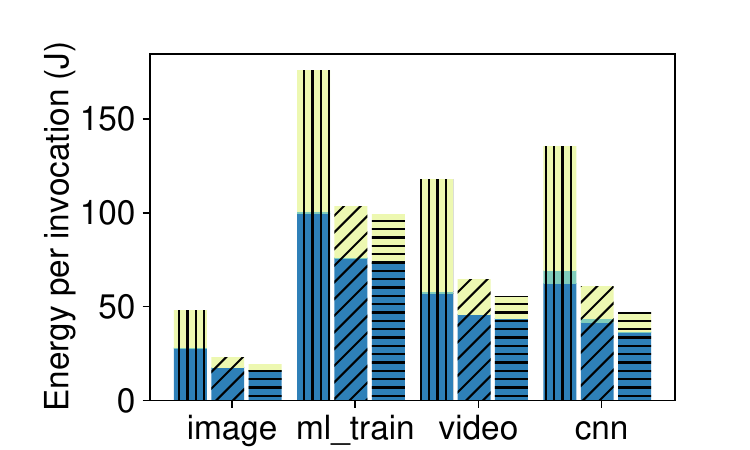}}
  \subfloat[Desktop-Small \label{subfig:desktop-small}] 
{\includegraphics[width=0.24\textwidth]{./figs/fig_2_smart_small_funcs.pdf}}
  \subfloat[Desktop-Large \label{subfig:desktop-large}] 
{\includegraphics[width=0.24\textwidth]{./figs/fig_2_smart_large_funcs.pdf}}
  \vspace*{-6pt}
  \caption{Function energy per invocation, measured in isolation. Server load and concurrency levels significantly impact the footprints, making this an unreliable method for energy measurement.}
    \vspace*{-6pt}
  \label{fig:isolated-all}
\end{figure*}

\vspace*{-0.5cm}
\section{Challenges in Function-level Energy Measurement}
\vspace*{-0.3cm}
\label{sec:challenges}

FaasMeter introduces energy accounting and control into the FaaS control plane (Figure~\ref{fig:design}). 
A key prerequisite is accurately measuring the energy consumption of individual function invocations.
The function energy footprints are used for virtualizing power in the control plane, for tasks such as energy pricing and power capping.
In this section, we discuss the tradeoffs of two broad approaches to energy measurement, and their suitability for FaaS workloads. 

\vspace*{-0.4cm}
\subsection{Direct Attribution}

In this approach, the hardware power sensors are read periodically, and the power consumed in the sampling interval is attributed to the software components (such as processes and functions) running during that interval.
It is used by popular tools like Scaphandre~\cite{scaph}, which rely on high-frequency CPU power measurements using RAPL.
The fundamental challenge is attributing a single power reading to a large number of concurrently executing components (such as multiple processes).
For this power \emph{disaggregation}, the total power is often evenly distributed~\cite{mukhanov_alea_2017, babakol_calm_2020}. %
High sampling rates and accurate hardware power sensors are vital: smaller sampling intervals (few milliseconds) contain fewer concurrently running components, which makes the disaggregation feasible.
Thus the direct attribution approach uses RAPL sensors which can be read with high-frequency (100s of Hz).
The overhead of power profiling is also a concern with direct attribution: our evaluation shows that the popular Scaphandre tool can increase CPU and energy consumption by more than 5\%.

Compared to CPU power sensors, \emph{system-level} energy can only be reliably obtained at low frequency, resolution, accuracy, and has temporal skew. 
The nature of FaaS workloads compounds the fundamental challenges and these measurement errors. 
Functions can be very short lived (<1 s), and FaaS servers run hundreds of functions concurrently.
These issues are illustrated in Figure~\ref{fig:pow-ml-train}, which shows system and CPU power when a \emph{single} compute-intensive ML training function is run in a loop.  
The ``Server'' power is measured through the IPMI and inbuilt chassis-level power sensor, and the ``Desk(top)'' uses a plug-level power meter for full-system power. 
There is a maximum of one active invocation at a given time, and each ``dip'' in power corresponds to the gap between invocations. 
The system power on the server has poor resolution and has large jumps. %
There are also large synchronization differences between the system and RAPL power on both platforms. 
On the desktop with a more accurate power-meter, the resolution is higher and the function signatures are more discernible, but the system power's time-diffusion problem persists.
These issues are amplified in FaaS servers running large numbers of small concurrent functions. %

\vspace*{-0.4cm}
\subsection{Model-based Power Estimation}

\emph{Power models} of the application and hardware~\cite{schmitt_online_2019, colmant_next_2018} are commonly used in energy measurement. 
For example, power can be modeled as a function of CPU utilization, which can be estimated with hardware performance counters such as instructions retired, cache misses, etc.
Compared to power, \emph{performance} measurement can be done with high fidelity, since a wide variety of fine-grained metrics are provided by the hardware and virtualized by the operating system. 
Once a power model for the server (and workload) is built, it can be used to infer the power consumption of individual software components based on their resource consumption.

Existing profilers are not cognizant of function boundaries and execution lifecycle.
For example, FaaS control planes employ keep-alive techniques to reduce the cold-start overheads, and keep the container resident in memory between invocations.
The function's container only consumes CPU resources when the function is invoked, which results in a highly non-stationary resource consumption behavior. 
Existing power models work well for stationary workloads, but highly ``bursty'' and concurrent FaaS workloads results in inferior fidelity.
To illustrate, the output of state of the art process-based accounting tool, PowerAPI~\cite{fieni_smartwatts_2020} is shown in Figure~\ref{fig:pow-boundary}.
The server is running multiple concurrent invocations of a single function, representing the easiest disaggregation case.  
The PowerAPI energy estimate of the function's containers and the number of ``active'' function invocations are shown in the figure. 
When the function is not running, we should expect its container's power to be zero.
However, we can see from the figure that the predicted energy is not correlated with the number of function invocations, and has temporal skew, which makes accurate energy footprints difficult to obtain.

\vspace*{-0.4cm}
\subsection{Validation}

Empirical validation of power profiling also poses many fundamental and practical challenges. 
The predominant metric for evaluating the accuracy of power profilers is difference between the measured power and the \emph{total} predicted power (used in~\cite{jay_experimental_2023}).
This metric, which we shall call the \emph{total power error}, does not capture the accuracy of the power footprints of the \emph{individual} components (e.g., processes or functions).

Individual power footprints are sometimes validated using \emph{isolated} measurements. 
The different applications are run individually, and all the system power can be attributed to the application as the ``ground truth'' power consumption. 
The energy footprints (energy consumed by a function per invocation) for such isolated measurements are shown in Figure~\ref{fig:isolated-all}.
We show the average energy per invocation over a 10 minute period where the same function is invoked in a closed-loop. 
A major drawback of this approach is that is is not ``online'', and does not capture the function's energy footprint under realistic loads. 
The hardware power consumption is highly dependent on the system load and the power states, and thus increasing the system load by running more concurrent functions affects the footprints.
In the figure, we run 1, 4, and 8 concurrent invocations of each function, and we can see that the footprints reduce with load, as the shared and idle power is amortized across individual invocations.
Thus, measuring energy in isolation is not suitable as a validation approach for function-level profiling, and \emph{additional metrics and validation methodology} is needed for increased confidence in the fidelity of the output of power profilers.

\vspace*{-0.4cm}
\subsection{Shared Components}
Measuring the energy of function invocations alone is insufficient and does not provide complete accounting. 
The FaaS control plane also performs many actions on behalf of the functions and is a major shared resource with its own energy footprint, which must be carefully attributed to the individual functions. 
The sandboxing and management of functions imposes significant work on the control plane, which also increases their energy footprint~\cite{sharma_hotcarbon22}. 
The time spent by OpenWhisk for a single (warm) invocation can be up to 600 milliseconds per invocation~\cite{il76-hpdc23}. 
This is separate from the actual function execution time (i.e., the ``function context''), and is a significant fraction of the total time (and hence resource and energy) consumption of the function.

The control plane interposes on many aspects of function execution asynchronously (such as dealing with the OS virtualization layer, caching container state, etc.). 
This results in a \emph{fuzzy boundary} between the function execution and the control plane, and exacerbates the challenges in  system-level energy measurement described previously.
The boundary is also fuzzy in time: since the function's initialized sandboxed is usually kept warm in memory~\cite{faascache-asplos21}, this results in a function's memory-energy footprint outlasting the function execution. 
The control plane's eviction and container life-cycle management operations also consume CPU resources and energy. 
The potentially large footprint of shared resources such as the control plane raises new challenges in \emph{fair} attribution: \emph{How should we measure and divide the control plane energy among the functions?}

\vspace*{-0.3cm}
\section{FaasMeter Energy Profiling}
\vspace*{-0.2cm}
\label{sec:design}

FaasMeter addresses the above challenges to provide fine-grained power accounting and control in FaaS environments.
It is a server-level system which integrates with monitoring infrastructure and the FaaS control plane such as OpenWhisk or \sysname~\cite{fuerst2023iluvatar} (see Figure~\ref{fig:design}). 
Robustness to measurement noise and workload dynamics is our key design requirement and influences our power modeling. 
We use three broad categories of input: hardware power measurements; OS and hardware level metrics (such as process-level CPU utilization and CPU performance counters); and a trace of function executions (start and end time of each invocation).
The availability and resolution of input metrics can be highly non-uniform (i.e., some hardware sensors may not be available on all platforms). 
FaasMeter is thus flexible about input data availability, and can work with a small subset of coarse-grained metrics if necessary.

Our power modeling is deliberately simple to be generalizable and robust, and we prefer explainable linear models to more complex ``black box'' models such as deep neural networks. 
We combine both the direct attribution and model-based techniques, and leverage repeated function invocations for statistical disaggregation. 
We provide \emph{complete} energy accounting of system-wide power by using Shapley value principles of fair division.
FaasMeter provides a wide spectrum of per-function energy footprints with different shared-energy contributions, which are suitable for different tasks pertaining to energy accounting and pricing, capping and control, etc.

Our power profiling has three major components.
The input power and workload measurements are disaggregated using the statistical model (Section~\ref{sec:design:disagg}), which we augment with a CPU power model (Section~\ref{sec:design:cpu}) when RAPL is available to provide a finer-grained power profile. 
The footprints are continuously updated based on a Kalman filter approach that we adapt for FaaS workloads (Section~\ref{sec:design:kf}).
FaasMeter can provide full-spectrum energy profiles such that the energy of shared components is attributed to functions, for which we use fair-division principles of Shapley values (Section~\ref{sec:design:fair}), which makes FaasMeter suitable for energy pricing. 
Figure~\ref{fig:isolated-all} illustrates this energy spectrum: the function's total energy profile comprises of its ``individual'' contribution due to function execution, as well as its share of the control plane energy and the idle energy of the server.

\vspace*{-0.4cm}
\subsection{Statistical Power Disaggregation}
\label{sec:design:disagg}

We use the repeated invocation of functions for statistical power disaggregation among functions and shared resources such as the FaaS control plane. 
The key idea is time-based attribution: the total system power at various points in time can be attributed among all the functions that are executing in a time period of $\delta$.
We then collect $N$ such sequential samples (each over a small $\delta$ interval) for the various power and workload metrics. 
$M$ is the total number of unique functions running on the server. 
The power measurement ($W$) can be system-wide power, CPU power, or the ``rest'' of system power which is total system power minus the CPU power. 

The key parameter for disaggregation is the ``function contribution to power'', which is the matrix $C$ with $M$ columns and $N$ rows. 
We use function running times as proxy for the contribution: $C[j]$ is the total amount of time the function $j$ was running during the interval.
Another useful parameter is the number of invocations or activations of each function in the interval, stored in matrix $A$. 
The total number of functions the server runs, $M$, is large, and the number of active functions with non-zero entries in $A$ and $C$ is small. 

We use simple linear regression for estimating the per-function \emph{power} consumption $X$:
\begin{equation}
X_{\text{Full}} =   \min_X{CX-W}
  \label{eq:cxw-full} 
\end{equation}
This is the simplest case which does not consider control plane or idle power, and the power values ($X$) obtained are referred to as the \textbf{full} power. 
The per-invocation \emph{energy}, $J$ in the interval is obtained by multiplying the power with the average function latency $\tau$:
$J_{\text{full}} = X_{\text{full}} \tau$.
In some cases, subtracting the idle server power provides more meaningful footprints:  $X_{\text{No Idle}} = \min_X{CX  - (W - W_{\text{idle}})}$. 

The choice of the measurement interval $\delta$ has important tradeoffs.  
At small intervals ($\delta \sim 10\text{ms}$), only a few functions are active, which makes $C$ sparse. 
In the extreme case, only one function is active, and all system power can be attributed to it without any further disaggregation.
However, in practice, the noise in system-level power measurement increases with the sampling frequency and increases the error. 
Conversely, larger $\delta$ values yield lower variance in power measurements, but denser contributions matrices, which increases errors in the linear regression solution. 
We use $\delta=1 \text{second}$ by default. 

\noindent \textbf{Shared Principals.}
As described in the previous section, the FaaS control planes can also be a significant energy consumer, along with other shared principals like the OS.
We augment the above statistical disaggregation to also include these shared principals as additional columns in the contributions matrix $C$. %

Shared principals like the control plane and OS are always running, so unlike functions, we cannot simply use running-time as their ``contribution''.
Instead, we use their CPU utilization as an indicator of energy use. 
For the control plane, we measure the CPU\% of all its processes.
Multiplying this CPU\% by the time-interval $\delta$ gives the fraction of time the control plane was running. 
However, this underestimates the control plane overhead, since function executions don't necessarily consume 100\% CPU. 
We thus normalize the control plane's contribution by the system-wide CPU:
\begin{equation}
  c_{cp} = \dfrac{\text{control plane CPU\%}} {\text{system-wide CPU\%}} * \delta 
\label{eq:cp}
\end{equation}
This yields $x_\text{cp}$, the control plane power, which is then divided among all functions using the Shapley value fair share principles described later in this section. 
We can similarly account for other shared components like the OS, by using the kernel's CPU-time and applying similar normalization.

\vspace*{-0.4cm}
\subsection{Online Estimation with Kalman Filtering}
\label{sec:design:kf}

FaasMeter continuously updates the function power estimates $X$ based on new measurements.
We use a Kalman filter framework~\cite{welch1995introduction, meinhold1983understanding} and adapt it to FaaS energy monitoring. 
This provides a tunable and robust approach with strong convergence properties that can adapt to changes in function and workload characteristics. 
The high-level intuition is to combine the previous estimates $X_{i-1}$ with the new measurements $C_i, W_i$, and also account for the changes or variance in the new measurements (i.e., the process and measurement noise).

The outline of our Kalman filtering algorithm is presented in Figure~\ref{fig:kf-code}. 
The filtered estimate is then given by:
\begin{equation} 
  \hat{X}_i = \alpha \hat{X}_{i-1} + \beta U_i + K Z_i,
\end{equation}
where $\alpha, \beta$ are tunable parameters. 
$Z_i$ is the ``innovation'', and is the error we get if we use the previous estimate with the new measurements. 
$K$ is the Kalman gain, which is the main component influencing how the innovation is distributed among functions and how footprints evolve. 
Our intuition is that updates to function footprints should be based on two factors: i) the number of invocations in the interval ($A$), and ii) their historical latency variance ($\sigma(T)$).
For instance, functions not executed in the interval should see no changes in their footprint.
The latency variance is a factor because our footprints are proportional to function latency ($C$), and functions with higher latency variance should receive a smaller update.
The latency variance is cumulative and also updated in each step (not shown in the algorithm), and $\gamma$ is the third tunable parameter. 
Based on the sampling tradeoffs, the measurement noise is set proportional to $1/\delta$.
The process noise ($P$) is updated after the Kalman step.
For new functions without any estimates, we set $\alpha=0, \beta=1$, and $K=0$. 
The initial estimates $X_0$ are obtained using statistical disaggregation on a large initial time-step ($N_{\text{Init}} \sim 2 \text{minutes}$).
Optionally, estimates from previous profiling runs or other servers in the cluster can also be used as the initial value. 
The subsequent Kalman steps are performed over the time-step $N_K$ in the range of $1-2$ minutes.
The same sparsity tradeoffs apply: smaller time-steps result in more sparsity and frequent updates, but are impacted more by latency variance and measurement noise.

\begin{figure}
  \lstset{basicstyle=\footnotesize,breaklines=true}
  \begin{lstlisting}[frame=single, mathescape=true, language=Python, numbers=left]
$\sigma(T)$: variance of function latencies 
$A_i$: num fn invocations during interval $i$
$r \propto 1/delta$
    
def Kalman-step($X_{i-1}, C_i, W_i, P_{i-1}$):
  $U_i = \min_X (C_iX - W_i)$
  $Z_i = W_i - C_i \hat{X}_{i-1}$
  $P = \alpha P_{i-1} + \gamma \sigma(T) $
  $K = {P A_i^T}/{A_i P A_i^T + r}$
  $P_{i-1} = (1 - KA_i)P$
  return $\hat{X}_i = \alpha \hat{X}_{i-1} + \beta U_i + K Z_i$

\end{lstlisting}
\vspace*{-0.2cm}
\caption{FaasMeter adapts a Kalman filter approach forupdating function per-invocation power $X$ over time.}
\vspace*{-0.2cm}
\label{fig:kf-code}
\end{figure}

\vspace*{-0.4cm}
\subsection{CPU Power Modeling}
\label{sec:design:cpu}

The statistical disaggregation technique described above has many advantages: it is simple, and requires only coarse grained power and latency measurements.
We combine this phenomenological approach with more a causal CPU power model for increased accuracy.
We build on the plethora of CPU power models~\cite{colmant_next_2018} and use hardware performance counters to map function CPU usage to power consumption.
A CPU model $\theta_{\text{CPU}}$ is built and used to predict the function's CPU-only power $X_{\text{CPU}}$ over each time-step.

$X_{\text{CPU}} = \theta_{\text{CPU}}(S)$, where $S$ is a vector comprising of the function's performance counters, normalized by the system-wide counters.
Our approach is similar to PowerAPI and SmartWatts~\cite{fieni_smartwatts_2020}, and uses the standard performance counters: \texttt{UNHALTED\_CORE\_CYCLES, UNHALTED\_REFERENCE\_CYCLES, LLC\_MISSES}, and \texttt{INSTRUCTION\_RETIRED}.
We use \texttt{perf} to obtain the function-container counters and aggregate the values for multiple concurrent containers of the same function. 
The model $\theta_{\text{CPU}}$ is trained using SVR with a linear kernel during initial operation~\cite{sklearnsvmsvr}. 
This model is stable as long as the function execution footprint (CPU work done and IPC) is stationary across invocations.
We continuously monitor the model error (difference between observed CPU power and the sum of all predicted function powers), and retrain the model if error exceeds a set threshold (default of 5\%).

We can combine the CPU power and rest of the system power estimates. 
The rest of the system power is obtained using the statistical disaggregation (and Kalman Filter): $X_{\text{Rest}} = \min_X (CX - W_{\text{Rest}})$, where $W_{\text{Rest}} = W_{\text{Sys}} - W_{\text{CPU}}$. 
When RAPL is available, FaasMeter's default is this \textbf{combined} mode, $X = X_{\text{CPU}} + X_{\text{Rest}}$.

\vspace*{-0.4cm}
\subsection{Fairly Attributing Shared Power}
\label{sec:design:fair}

FaasMeter provides fine-grained full-spectrum function energy profiles by using fair division and Shapley value principles~\cite{winter2002shapley} to divide shared energy among functions. 
The Shapley value of each function would be it's ``true'' energy footprint, and satisfy many desirable properties, and can be considered the gold-standard of energy attribution~\cite{dong_rethink_2014}. 
Unfortunately, computing Shapley values requires sampling an exponential number of energy readings that cover all the permutations of function invocations (i.e., entries of the $C$ matrix), and the true marginal energies for all function combinations, which we have no way of obtaining accurately. 
Exact Shapley values are thus infeasible and impractical, especially considering measurement noise and under an online setting.
Instead, we approximate the Shapley values by satisfying its four properties in a best effort manner:
\begin{enumerate}[wide,labelwidth=!,labelindent=0pt,topsep=0pt,itemsep=-1ex, partopsep=1ex,parsep=1ex]
\item \emph{Efficiency:} the sum of all function footprints should add up to the total system-level energy. We try to achieve this by minimizing net error in the Kalman filter. 
\item \emph{Null-player} property requires functions not executing to have 0 energy, which we get by construction of our $C$ matrix. 
\item \emph{Symmetry:} identical functions (both in their code and invocation frequency) should have the same footprints.
\item \emph{Linearity:} the total shared-resource energy attributed to a function should be the sum of the individual shared resources. 
\end{enumerate}

Based on these principles, we divide the shared idle and control plane energy among the functions.
Idle power is a ``static'' shared resource, and it must be evenly among all functions (proof omitted, but follows similarly from~\cite{islam2016new} which applies Shapley values to datacenter power). 
On the other hand, the control plane energy is ``dynamic'' and depends on its use (i.e., how many functions are invoked), and it must be divided among functions on a \emph{per-invocation} basis. 
The efficiency and linearity properties require that we add the individual and static and dynamic energy shares to obtain the total energy for each function:   \vspace*{-0.2cm}
\begin{equation}
  \label{eq:J_total}
  J_{\text{total}}  = J_{\text{indiv}} +   \phi_{\text{cp}} +   \phi_{\text{idle}},
\end{equation}   
where $\phi_{\text{cp}}, \phi_{\text{idle}}$ are the shares of control plane and idle energy respectively.
$J_{\text{indiv}}$ is obtained by discounting the idle power, i.e., using $X_{\text{No Idle}}$. 
$ \phi_{\text{cp}} = J_{\text{cp}}A_i/\sum(A) $, where $A_i$ is the number of invocations of function $i$, and $A$ is the vector of $A_i$. 
We divide the control plane energy proportional to function invocation frequency (over the time-step). 
$\phi_{\text{idle}} = J_{\text{idle}}/M$, where $M$ is the number of unique active functions in the interval, which is the number of non-zero entries in $A$. 
The function footprints are then the \emph{per-invocation} energy, which is $J_{\text{total}}/A $. 

The above total footprint  gives applications a full and complete picture of their energy/carbon consumption.
``Charging'' functions for resources they do not explicitly consume such as server's idle energy consumption may seem excessive and unfair, and thus by providing separating the energy components, FaasMeter provides the ability to create new energy accounting and pricing schemes. 
For instance, when developers are optimizing the energy footprint of their functions, only the direct and individual energy (without any control plane or idle overheads) is suitable.
For energy pricing, the full footprint $J_{\text{total}}$ may be used, which provides incentives for FaaS providers to reduce their idle energy overheads by increasing utilization and other efficiency improvements.

\vspace*{-6pt}
\section{Implementation and Validation Methods}
\vspace*{-6pt}
\label{sec:impl}

FaasMeter is integrated with and extends the \sysname~FaaS control plane with power accounting and virtualization. 
It is implemented in Python and Rust in around 6,000 lines of code. 
Our new energy-monitoring layer in \sysname~provides time-series of power and other metrics to the energy profiler, which is a Python module. 

Different full-system power sources are supported.
For servers equipped with chassis-level current and power sensors, we use IPMI to query the BMC (baseboard management controllers).
We also support external plug-level power meters, and query their power via serial or telnet interfaces.
Many low-power edge devices also provide system-level power.
For example, the Nvidia Jetson Orin AGX has current and power sensors which we query using tegrastats.
Finally, for laptops and other battery powered devices, the battery charge controller can provide the energy discharge, which we can obtain via ACPI interfaces, and obtain the power consumption.
We use the \texttt{perf} tool for both the system-wide RAPL and per-function RAPL and per-function CPU performance counters. 

\noindent \textbf{Software Power Capping.}
FaasMeter provides the various energy footprints via an RPC interface.
These are used for various energy accounting and management tasks such as pricing, resource allocation, and power control.
We add energy as a new resource dimension in \sysname~for resource management. 
We have implemented a system-wide power capping mechanism, which allows setting upper-limits on the server-level power consumption. 
This power capping is integrated with the existing \sysname~resource scheduling architecture, and can be combined with various scheduling algorithms. 
We use the FaasMeter footprint of the function at the head of the queue, $J_{\lambda}$, to estimate the increase in energy consumption due to running the function $\lambda$. 
The total estimated energy consumption over an interval of $t$ seconds is then $Wt + J_{\lambda}$, which must be less than the $W_{\text{cap}}t$.
Incorporating the function footprint reduces the power cap violations, since it provides a better prediction of the power consumption. 
Without energy footprints, we use a large buffer $b$, such that $W +b < W_{\text{cap}}$, but this increases the wait time, especially for small functions.

\noindent \textbf{Power De-noising and Synchronization.}
We filter and synchronize the raw power signals before using them for disaggregation. 
``External'' power using external plug-level power meters and even BMC/IPMI can have a time-skew in their measurement and reporting path. 
Synchronizing the system-power is crucial---otherwise energy is attributed to previous/future functions, reducing the footprint accuracy. 
We correct the temporal skew (i.e., lag) by correlating (in time) the power signal with some other reference. 

\begin{figure}
  \centering
  \includegraphics[width=0.3\textwidth]{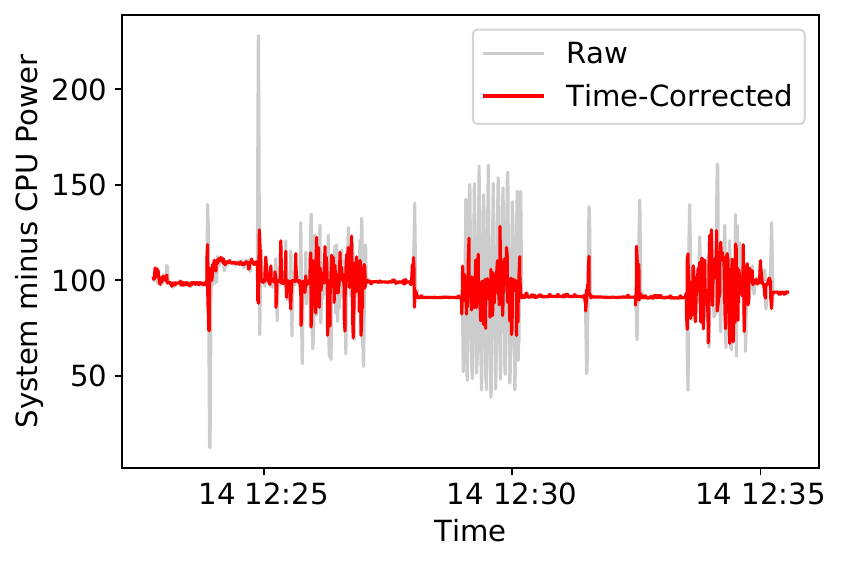}
  \vspace*{-5pt}
  \caption{Synchronization of system-level power.}
  \label{fig:pow-noise}
    \vspace*{-5pt}
\end{figure}

For instance, we have observed that IPMI power has significant lag, when compared to the RAPL power which is much more ``real-time''. 
This can be observed from Figure~\ref{fig:pow-noise} which shows the difference between system-level and CPU power.
The workload is a CPU dominant application (ML training), and the server has no other major dynamic power-consuming devices (no GPU etc.).
Thus we should expect this difference to be constant.
However, we see that the raw difference shows significant variance, which we attribute to the measurement lag in the IPMI power-sensor. 

We find and fix this lag using simple signal-shifting methods~\cite{Pearson2019}.
We find the time-offset $s$ which minimizes the chi-square difference between the power signal, $W$ and the reference signal $R$, after normalizing by the average:
\begin{equation}
  \label{eq:pow-correct}
s^* =  \min_{s} \left(\frac{W(t+s)}{\bar{W}} - \frac{R(t)}{\bar{R}}\right)^2
\end{equation}
This can be solved using general optimization solvers such as Limited-memory BFGS~\cite{scikit-learn}, after setting bounds on $s$ of a few seconds. 
The difference in the power and reference signal after time-skew correction is also shown in Figure~\ref{fig:pow-noise}, and we can see a significant reduction in the variance (i.e., the noise). 
FaasMeter computes this skew  both during an initialization phase, and periodically, to capture any drift. 
The reference signal is CPU power by default---other reliable load metrics like CPU instruction and cycle counters are the fall-back synchronizing inputs.

\vspace*{-0.4cm}
\subsection{Validation Methods and Metrics} %
\label{sec:impl:validation}

Our methodology and metrics required for validating energy footprints in highly dynamic and heterogenous FaaS environments fall into two broad classes. 
\emph{External validation} entails comparing the FaasMeter energy footprints with other reliable energy measurement methods, and provides the primary accuracy metrics.
On the other hand, the \emph{internal validity} looks at the consistency of energy footprints with respect to each other, or other system utilization and performance metrics. 

External validity for energy disaggregation is challenging: we want to estimate the function's energy contribution in a long and dynamic workload.
Our primary benchmark and ``ground truth'' for external validity is the \textbf{marginal energy}, which we compute by running two nearly identical workload traces, and subtracting their \emph{total} energy consumption. 
A workload trace ($\mathcal{T}$) is characterized by the set of functions ($\mathcal{S}$), their IAT CDFs, and the total duration. 
Even using extremely coarse-grained power measurements, we can obtain $\mathcal{J}(\mathcal{T})$, the total energy consumption of running the workload trace on a server.  
The marginal energy of a function $f$ is obtained by running a new trace $\mathcal{T}(\mathcal{S} - f)$ which doesnt contain the function, and is given by:
\begin{equation} \vspace*{-6pt}
  \mathcal{M}_f = \dfrac{\mathcal{J}(\mathcal{T}(\mathcal{S})) - \mathcal{J}(\mathcal{T}(\mathcal{S} - f))}{\text{number of invocations of f in S}}.
  \end{equation}
  The marginal energy is thus the increase in total energy consumption caused by the function. 
Note that it \emph{does not} account for the idle server energy, since it is present in both the traces.

This marginal energy validation is essential, since energy is highly sensitive to the system power states (such as CPU frequency).
Measuring the function energy footprints ``in isolation'' is a simpler and alternative technique, where a single function is run without any other function, and per-invocation energy is obtained from the total system energy.
But because of the different power states, the per-function energy footprints obtained through this conventional and simpler method have a very high range, dependent on the system load. 
Figure~\ref{fig:isolated-all} shows the energy per invocation with the isolated measurement technique when different numbers of the same function are executed concurrently, and we can see significant differences (more than $10\times$) in footprints based on the load. 

\begin{table}
  \begin{tabular}{|l|l|}
    \hline 
    Metric & Definition \\
    \hline
    Individual-Difference & ${|J-J^*|}/{J^*}$ \\
    \hline 
    Cosine-Similarity & ${J \cdot J^*}/{||J|| ||J^*||}$ \\
    \hline 
    Total-Error & $E[\ |W(t) - \hat{W}(t)|/W(t)\ ]$  \\
    \hline 
    Latency-norm.-Variance & $ {\sigma(J)}/{\sigma(T)}$ \\
    \hline 
  \end{tabular}
  \vspace*{-5pt}
  \caption{External and internal validity metrics. $J$ and $\hat{W}$ are profiler outputs, $J^*$ is ground truth, and $T$ is function latency.}
  \label{tab:metrics}
  \vspace*{-10pt}
  \end{table}

\noindent \textbf{Validation Metrics.}
Table~\ref{tab:metrics} lists our  external and internal validation metrics. 
For external validity, we define and compute different \emph{distance metrics}  between the energy profiler output vector ($J$) and some reference ground-truth $J^*$ (e.g., marginal energy footprints).
The first metric provides the \emph{per-function} individual difference to the ground truth.
Since the marginal per-invocation energy is obtained by running a separate, smaller workload, it may not always reflect the ``live'' online energy.
These are primarily due to differences in system power consumption and efficiency, and underlying hardware control.
Deviations from the marginal can also arise because of different attribution policies for idle and other shared energy. 
To achieve ``complete'' energy accounting, the footprints may be elevated for \emph{all} functions. 
To account for these issues, we use the \emph{cosine similarity} between the energy footprints, to capture the ratios of energy footprints among the different functions.
Higher cosine similarity (closer to 1) indicates that a closer footprint match.
\emph{Cosine similarity with the marginal energies is our primary external validation metric.}

The second category of metrics are for internal validity.
The conventional and popular metric for energy profilers captures the ``completeness'' of accounting, by computing the difference over time,  between the observed ($W(t)$) and predicted total power ($\hat{W}(t)$).
Optimizing solely for this total error can be at the expense of the error in energy footprints. 
This total power error is thus of secondary importance to FaasMeter, and is controllable via our Kalman filter parameters: higher values of $\beta$ in Figure~\ref{fig:kf-code} reduce the total error but result in higher variance in footprints.

Our primary internal validation metric is the variance in the energy footprints, normalized to the variance in latency. 
This helps use evaluate the precision and feasibility of FaasMeter footprints for energy pricing. 
Currently, cloud functions are priced based on their execution time (i.e., latency), and thus the latency variance serves as the baseline for comparison.

\vspace*{-6pt}
\section{Experimental Evaluation}
\vspace*{-6pt}
\label{sec:eval}

The goal of our experimental evaluation is to validate FaasMeter's energy footprints in a diverse range of workload and hardware configurations, and investigate their effectiveness for energy control and pricing use cases.
We evaluate FaasMeter's energy footprints on more than 100 workload traces (including marginal energy traces) using seven external and internal validity metrics.
These workload and measurement traces and associated ground truths will be made publicly  available, to help build better energy profiling models and accelerate empirical sustainability research. 
We also present its effectiveness in power-capping and energy pricing.

\noindent \textbf{Measurement Platforms.}
Although FaasMeter's core ideas can be generally used across different FaaS control planes, we use \sysname~for lower-variance and reliable experimentation.
We use three different types of hardware platforms:
\begin{enumerate}[wide,labelwidth=!,labelindent=0pt,topsep=0pt,itemsep=-1ex,partopsep=1ex,parsep=1ex]
\item \emph{Server:} Supermicro X11DPT-PS board with 2 48-core Intel Xeon Platinum 8160 CPUs and 1TB of RAM, running Ubuntu 20.04. It idles at 95 Watts. %
\item \emph{Desktop:} Dell Optiplex with 12th Gen Intel i5-12500 running Ubuntu 20.04.5. It has idle power of 15W. Power is being measured using an external SpecPower-approved power-meter (Instek GPM-8310) every 0.25 seconds through the meter's telnet/SCPI interface. 
\item \emph{Edge:} Nvidia Jetson Orin AGX~\cite{jetson_2022}. Power is measured using inbuilt current sensors (via \texttt{tegrastat}) and also validated with an external power meter. We run a combination of CPU and GPU functions on this edge device. 
\end{enumerate}

\noindent \textbf{FaaS workloads} are generated to be representative samples of the Azure function trace~\cite{shahrad_serverless_2020}, with its functions mapped to the nearest functionbench functions~\cite{kim_functionbench_2019} based on their execution time. 
We use 30 minute workload traces with different functions and inter-arrival-time (IAT) distributions. 
We also use other non-stationary and bursty workloads as appropriate.
We use a sufficient keep-alive cache~\cite{faascache-asplos21} to ensure $>99\%$ warm starts, and all the latency and energy footprints we report are for warm starts. 
Cold-starts are tagged by the FaaS control plane, and we can obtain separate cold and warm energy fingerprints if necessary. 
We use different sizes and types of functions for our measurements---the function characteristics are described in Table~\ref{tab:fns}.

\begin{table}
  \resizebox{0.45\textwidth}{!}{%
  \begin{tabular}{|l|r|l|}
    \hline
    Name & Latency (s) & Description \\
    \hline
    dd & 0.7 & Read and write local storage. \\
    image & 1.5 & Performs several transformations on an image. \\
    video & 7.8  & Download and grayscale a small video. \\
    AES & 1.4 & Encrypt and decrypt payload multiple times. \\
    json & 0.25 & Download, parse, and serialize a json blob. \\
    CNN & 1.3 & Inference on a TensorFlow model (CPU and GPU). \\ 
    ml\_train & 5.1 & Train a regression model on a 20 MB dataset \\
    \hline 
  \end{tabular}
  }
  \vspace*{-3pt}
  \caption{Functions used in our empirical energy analysis. Latency is average warm running time on the desktop.}
  \label{tab:fns}
  \vspace*{-3pt}
\end{table}

\begin{figure*}
  \centering
  \vspace*{-4pt}
  \includegraphics[width=1.0\textwidth]{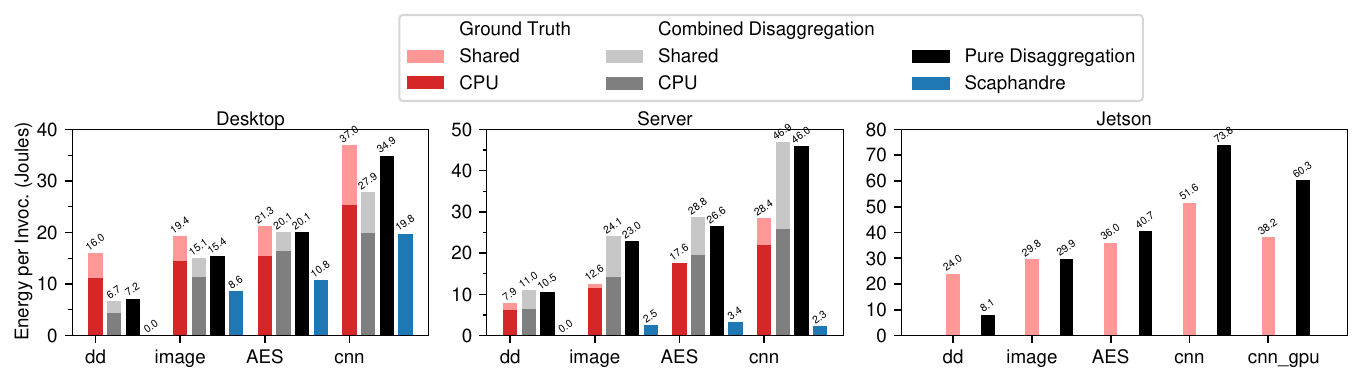}
  \vspace*{-18pt}
   \caption{Energy-per-invocation for a four-function trace. Marginal energy serves as ground-truth. FaasMeter's footprints with combined and pure disaggregation are accurate across all three hardware platforms. Scaphandre~\cite{scaph} provides inaccurate footprints, especially for non-CPU-intensive functions (dd), and requires x86 RAPL counters (not available on Jetson).}
   \label{fig:accuracy-all}
     \vspace*{-4pt}
 \end{figure*}

 We use the Kalman filter with $\alpha=0.8, \beta=0.2, \gamma=0.1$, with an initial window of 100 seconds, and subsequent intervals of 60 seconds. 
We compare against Scaphandre~\cite{scaph}, a popular process-level CPU-only power profiler which uses the direct disaggregation approach.
We have implemented additional post-processing for Scaphandre to turn the process-level power output to per-function-invocation metrics, by correlating the host process id with the function (\sysname~runs functions in \texttt{containerd} containers by default).

\vspace*{-6pt}
\subsection{External Validity vs. Marginal Energy}
\label{sec:eval:fm}
\vspace*{-4pt}

We focus the first part of our evaluation on one heterogeneous trace with four functions.
The function IATs are scaled for the three hardware platforms, such that the desktop and Jetson are at 100\% load, and the server is at 50\%.
The per-invocation energy footprints for each function are shown in Figure~\ref{fig:accuracy-all}.  
We compare against the marginal energy as the ground truth baseline, and also show the results of Scaphandre~\cite{scaph}. 
We evaluate two different FaasMeter configurations.
In the \emph{combined} mode, the CPU power is measured separately and added to the rest of the system power using statistical disaggregation. 
In the \emph{pure disaggregation} mode, only the coarse grained statistical disaggregation is used on the full-system power.
In both cases, we subtract the idle hardware power, and thus compute the $X_{\text{No idle}}$ described in Section~\ref{sec:design}. 

\begin{table}
  \begin{tabular}{|l|l|l|l|}
    \hline
  Platform & Full Disagg. & Combined Disagg. & Scaphandre \\ \hline
  Desktop & \textbf{0.985} & 0.984 & 0.910 \\ \hline
  Server & \textbf{0.998} & 0.998 & 0.623 \\ \hline
  Jetson & \textbf{0.992} & N/A & N/A \\ \hline
  \end{tabular}%
  \vspace*{-6pt}
  \caption{The high cosine similarity of footprints indicates high accuracy of FaasMeter with respect to ground truth.}
    \vspace*{-6pt}
  \label{tab:errors_cosine}
\end{table}

On the desktop, both these approaches are within 1--40\% of the marginal energy (the Individual-Difference metric in Table~\ref{tab:metrics}). 
Scaphandre measures only CPU power, and is unable to attribute energy the disk-intensive \texttt{dd} function, and has a 100-130\% difference vs. marginal for the rest of the functions on the desktop. 
Since the marginal energy is measured as the difference of the workloads and does not capture idle energy, it may have a higher absolute Individual-Difference.
For instance, the server has a high idle energy and only a 50\% load, which causes FaasMeter footprints to be consistently and proportionally larger than the marginal energies. 
Our cosine similarity metric captures this proportionality, and is shown for this workload in Table~\ref{tab:errors_cosine}. 
\emph{FaasMeter's footprints are within 98.4\% and 99.8\% of the ground truth, providing evidence of FaasMeter's accuracy.}

\begin{figure}
\subfloat[Function latency variance is uncorrelated with energy footprint error.\label{fig:err-varT}]
{\includegraphics[width=0.26\textwidth]{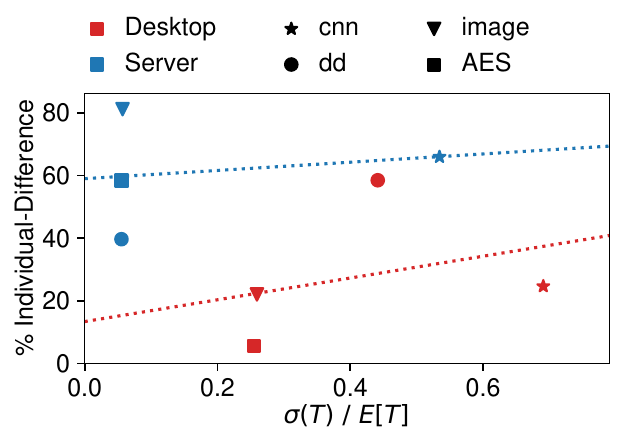}}
  \subfloat[Energy of identical functions can be clustered.\label{fig:J-clustering}]
  {\includegraphics[width=0.22\textwidth]{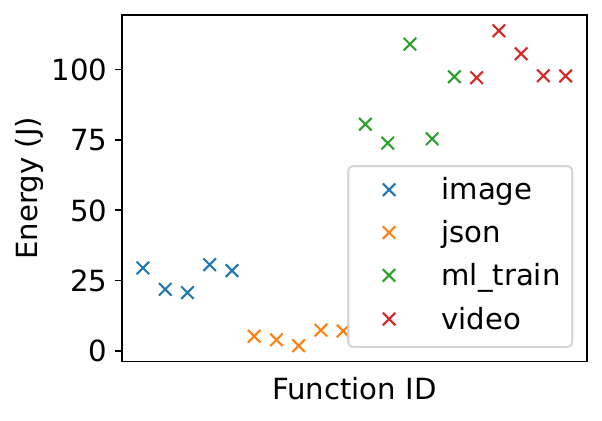}}
  \vspace*{-5pt}
  \caption{Variance in latency is primary source of difference vs. the marginal energy. FaasMeter provides the symmetry property for functions.}
  \label{fig:efficiency-symmetry}
  \vspace*{-9pt}
\end{figure}

FaasMeter's accuracy is a result of its statistical disaggregation approach.
The primary source of large Individual-Difference (high error) is the variance in function latency, since our model uses it for both power and energy prediction.
However, FaasMeter is robust to such latency variance, as seen in Figure~\ref{fig:err-varT}, which shows a lack of any significant correlation. 
On the server with 50\% load, the coefficient of variation of latency is low, but the high idle energy results in the larger offset vs. marginal.
The desktop has a 100\% load and higher latency variance, but is closer to the marginal energy as a result.
\emph{Thus, even with variation in latency due to differences in function input, contention, etc., FaasMeter can provide accurate energy footprints.}

On the Jetson platform, we do not have access to a CPU power model---but the pure disaggregation approach still provides a high 99.2\% cosine similarity.
FaasMeter is also able to measure the GPU function's footprint---highlighting its effectiveness on heterogeneous platforms without specialized power measurement instrumentation. 
Scaphandre uses the x86 RAPL counters, and is unable to work on the ARM Jetson platform. 
Its process-level direct attribution is unable to account for all the CPU power on the x86 server, and has an error of $10\times-23\times$. 
Moreover, it has a high profiler overhead due to periodically scanning process info via \texttt{procfs}: its CPU consumption on the server is more than 5\%, causing a 15 Watt increase in power consumption.
For comparison, the \emph{combined} CPU consumption of FaasMeter and \sysname~is 3\%.
Scaphandre's high error on the server is also due to its high profiling overhead: the high latency of synchronously reading and disaggregating \texttt{procfs} information for more than 1000 processes (corresponding to function active and kept-alive containers) results in highly stale RAPL readings (several seconds) and inaccurate disaggregation.

\emph{Using noisy system-level power is one of FaasMeter's main features, and these results illustrate its robustness and accuracy. It provides more than an order of magnitude accuracy compared to existing process and CPU based tools, and is effective even on heterogeneous hardware.}

\begin{figure*}
    \subfloat[Bursty function invocations.\label{fig:stacked-bursty}]
    {\includegraphics[width=0.31\textwidth]{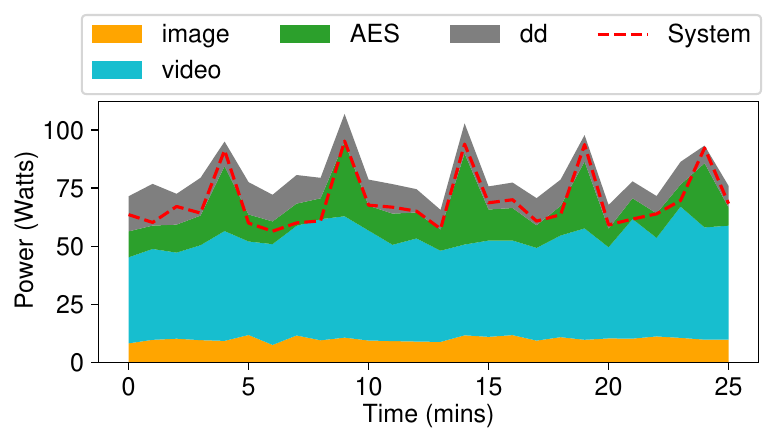}}
    \subfloat[New functions being added.\label{fig:stacked-morefuncs}]
    {\includegraphics[width=0.31\textwidth]{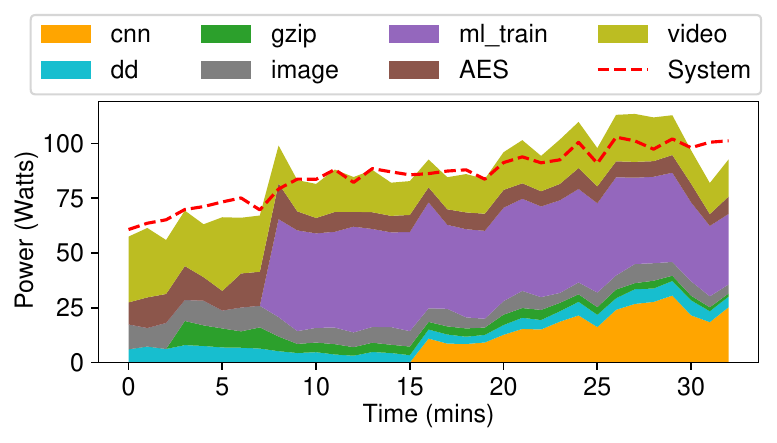}}
    \subfloat[Total power error. \label{fig:pow-err-all}]
    {\includegraphics[width=0.28\textwidth]{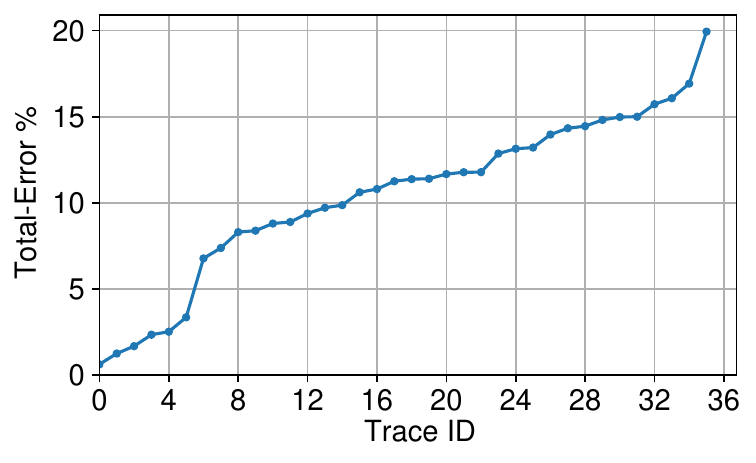}}
    \vspace*{-6pt}
    \caption{The misprediction of total power is small, even for dynamic and non-stationary workloads.}
    \label{fig:stacked-all}
    \vspace*{-3pt}
\end{figure*}

\vspace*{-6pt}
\subsection{Internal Validity}

\textbf{Symmetry}, a Shapley value property, requires that identical functions have similar energy footprints. 
We run a large number (20) of different functions, with each function belonging to the one of four classes (\texttt{image, json, ml\_train, video}). 
For FaasMeter, these are 20 different functions, and their footprints are shown in Figure~\ref{fig:J-clustering}. 
We can see that the functions can be clustered based on their energy footprints, i.e., the functions running image processing have similar per-invocation energy, thus providing the symmetry requirement.

Next, we look at the \textbf{Total-Error} from Table~\ref{tab:metrics}, which measures the difference in measured and the estimated total power (which is the aggregation of individual estimated footprints).
This is also the \emph{efficiency} property of Shapley values: we want all the energy accounted for.
The stacked energy footprints of different workloads are shown in Figure~\ref{fig:stacked-all}. 
It also illustrates a common use-case for profiling tools, which is to identify ``top'' energy consumers and their relative contributions. 
Figure~\ref{fig:stacked-bursty} shows the energy contribution of four functions in a ``bursty'' workload on the desktop, and we can see that the Kalman filter is able to track and estimate the total power. 
Another challenging workload is when the functions are dynamically introduced into the workload, and the ``active set'' is dynamic. 
The total energy breakdown of this workload is shown in Figure~\ref{fig:stacked-morefuncs}, where we also see low Total Error.
As mentioned in Section~\ref{sec:impl:validation}, minimizing Total-Error is \emph{not} our primary objective, since it can often reduce the accuracy (cosine similarity and other external validation metrics). 
FaasMeter's Total-Error across 35 workloads (with different functions and IATs) on the three hardware platforms is shown in Figure~\ref{fig:pow-err-all}. 
\emph{Because of the Kalman filter and continuous footprint refinement, we see that the Total-Error is small, and less than 10\% for more than 50\% of the tested workload configurations.}

\begin{figure}[!h]
  \vspace*{-6pt}
\includegraphics[width=0.48\textwidth]{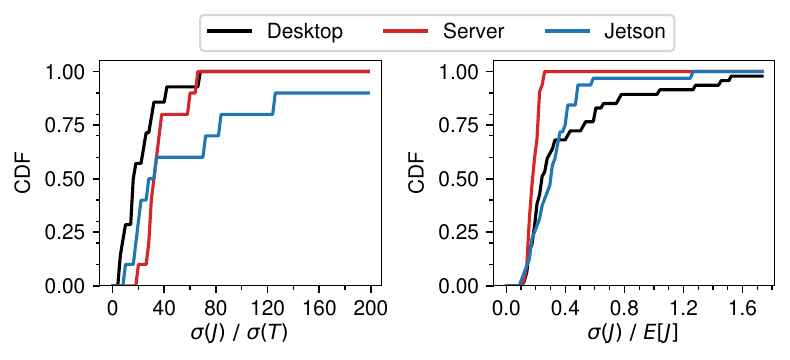}
  \vspace*{-20pt}
  \caption{Variance of FaasMeter energy footprints is low, making energy-based pricing feasible.}
  \label{fig:jpt-all}
  \vspace*{-10pt}
\end{figure}

For energy-pricing, we want the footprints to be ``stable'' and have low variance.
Figure~\ref{fig:jpt-all} (right) shows the coefficient of variation ($CoV = \sigma(J)/E[J]$) of FaasMeter for more than 50 workload traces.
The CoV depends on the measurement noise of the underlying hardware platform and the workload, and is akin to FaasMeter's \emph{precision.} 
The CoV is less than 0.3 on all three platforms for 60\% of the traces, indicating feasibility of using energy footprints as an accounting and pricing measure.
As noted before, the desktop energy variance is higher because its workloads are run at near-100\% load. 

Finally, we look at the \textbf{latency-normalized-variance} metric from Table~\ref{tab:metrics}.
This metric is another proxy for the stability of the energy pricing, since it compares against variance in currently used running-time based prices. 
Figure~\ref{fig:jpt-all} (left) shows the CDF of the average normalized energy variance across all functions, for more than 50 workload traces across the three platforms.
This ratio is less than 40 for more than 90\% of the desktop and server workloads. 
Note that \sysname~provides extremely low variance in latency which is $50\times-100\times$ lower than OpenWhisk~\cite{il76-hpdc23}.
The variance in latency and pricing is also significant in public FaaS clouds~\cite{ustiugov_analyzing_2021, schirmer_night_2023}. 
Thus, if were to use FaasMeter footprints with the more widely used OpenWhisk, we achieve a latency-normalized-variance ratio of close to $1$, indicating that the variance in energy price would be similar to the current variance in latency-based pricing. 
Thus, \emph{FaasMeter energy footprints have high precision (i.e., low variance) and may be used for energy pricing.}

\vspace*{-6pt}
\subsection{FaaS Energy Management}
\begin{figure}
      \subfloat[Power and workload dynamics. \label{fig:soft-cap:exceed}]
    {  \includegraphics[width=0.25\textwidth]{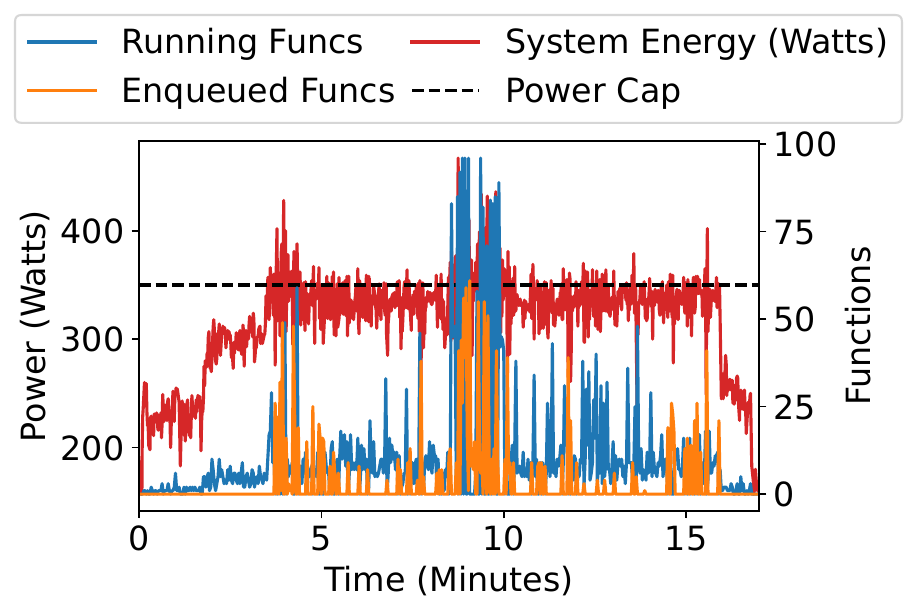}}
   \subfloat[Function latencies. \label{fig:soft-cap:lat}]
   {  \includegraphics[width=0.2\textwidth]{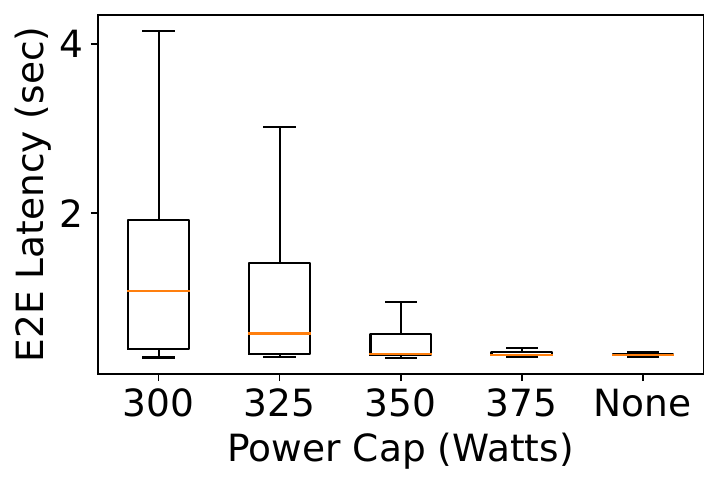}}
       \vspace*{-5pt}
   \caption{Impact of software power capping on system energy and function invocation latency.}
   \label{fig:ipmi}
    \vspace*{-6pt}
\end{figure}

FaasMeter's \textbf{power capping} can be used in environments without hardware capping support, and integrates with existing FaaS resource management.
Power capping affects the function latency by increasing queue wait times. 
This behavior can be seen in Figure~\ref{fig:soft-cap:exceed} where queueing increases when there are sustained periods of high energy usage across the entire system. 
The figure shows a workload trace with non-stationary IATs. 
The ``overshoot'' of our software capping approach is small, and less than 3\%. 
Figure~\ref{fig:soft-cap:lat} shows the function latency for different power caps on the server: reducing the power-cap predictably increases the latency mean and variance.
\emph{Software power capping is a viable mechanism for fine-grained FaaS scheduling to manage power usage.}

\begin{figure}
  \centering
  \includegraphics[width=0.45\textwidth]{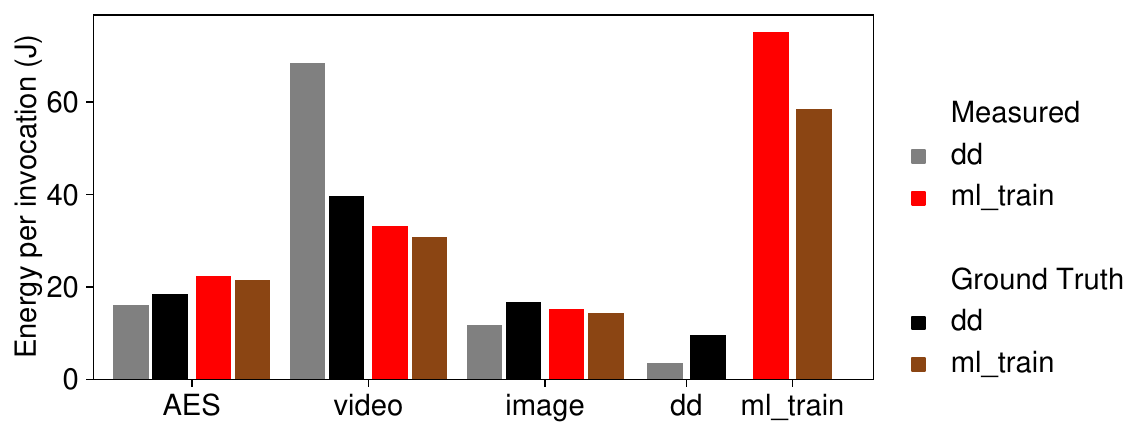}
  \vspace*{-5pt}
  \caption{The three functions are co-located either with \texttt{dd} or \texttt{ml\_train}. This choice has negligible impact on both FaasMeter and ground-truth footprints.}
  \label{fig:noisy}
  \vspace*{-6pt}
\end{figure}

\noindent \textbf{Noisy Neighbors.}
\emph{Are function energy footprints impacted by co-located functions?}
To answer this, we run functions with different ``neighbors''.
Three functions (\texttt{image, AES, and video}) are run together either with \texttt{dd} or \texttt{ml\_train} as the co-located function in Figure~\ref{fig:noisy}. 
The marginal energy in the two cases are nearly identical, and differ by at most 5\%.
The FaasMeter footprints are also similar, and vary by roughly 5--10\% between the two cases. 
This independence of energy footprints is important for energy modeling and optimization, since we can use linear models for total system energy. 
\emph{Function footprints are not dependent on their neighbors, and FaasMeter can disaggregate energy for workloads with small differences, further validating its accuracy.}

\vspace*{-8pt}
\section{Related Work}
\vspace*{-8pt}
\label{sec:related}

Our work is inspired and motivated by the quest to make energy as the first-class resource~\cite{flinn_powerscope_1999} in many environments.

\noindent \textbf{Sustainable Computing.}
The challenge of reducing the carbon footprint of cloud systems and applications has led to ``carbon-first'' system designs~\cite{souza2023ecovisor, anderson_treehouse_2022}. 
The operational carbon footprint is computed by multiplying the energy consumption and the grid carbon intensity~\cite{maji2022carboncast, watttime}, which has led to new energy tracking tools~\cite{cncf_kepler}, but CPU power profiling continues to dominate. 
~\cite{he_energat_2023} introduces fine-grained NUMA-aware CPU energy measurement for individual applications. 
Tools for tracking the carbon footprint of AI applications mostly focus on large ML training batch jobs without multi-tenancy~\cite{henderson_towards_2020, anthony_carbontracker_2020}.
The prominence of energy as the primary resource in cloud computing has led to many proposals for energy-pricing~\cite{narayan2013power}. 
The universal nature of energy and the fine-grained nature of functions makes it appealing to use in conjunction with the resource-as-a-service abstraction~\cite{agmon2014rise}.

\noindent \textbf{Energy Control.}
In the context of FaaS,~\cite{rastegar_enex_2023} presents DAG scheduling for functions with a purely CPU-model based approach, but without any empirical power measurement or validation.
DVFaaS~\cite{tzenetopoulos_dvfaas_2023} implements PID control for CPU frequency for minimizing latency QoS violations for function chains. 
More generally, a combination of hardware and software techniques for energy capping can be effective~\cite{zhang_maximizing_2016}.
Due to hardware heterogeneity, we use a purely software approach for controlling system-wide power, and focus on empirical energy footprints of individual functions.

\noindent \textbf{Fair Attribution.}
The problem of fairly sharing the energy consumption of shared resources occurs in many environments such as VM hosting~\cite{islam2016new, jiang_virtual_2017} and datacenter cooling~\cite{jiang_non-it_2018, wang_real-time_2020}. 
Shapley values~\cite{vergara_sharing_2015} provide many desirable properties such as linearity and envy-freeness, and have also been used for energy accounting of mobile applications~\cite{dong_rethink_2014}.

\noindent \textbf{Mobile and embedded computing}~\cite{fonseca_quanto_2008} faces similar energy measurement challenges. 
Disentangling shared OS and hardware energy consumption for applications has been done through tracing requests across various contexts and carefully attributing async tasks~\cite{oliner2013carat, carroll2010analysis, pathak2012energy, yoon_appscope_2012}.
Along with the uncertainty of the control plane, the highly dynamic and non-stationary nature of function workloads, and high degree of multiplexing makes such tracing challenging for FaaS.

\vspace*{-10pt}
\section{Conclusion}
\vspace*{-8pt}
FaasMeter brings energy measurement, accounting, and control to serverless computing.
As a key prerequisite, we focus on energy profiling of functions.
Using statistical disaggregation, Kalman filtering, and Shapley value principles, we provide full-spectrum energy footprints with 99\% over  accuracy.
We provide external validation with marginal energy on different hardware platforms, and develop a new energy metrology framework for FaaS.
FaasMeter makes the first step towards power virtualization for FaaS, and implements energy control and pricing.

\noindent \textbf{Limitations and Future Work.}
FaasMeter only considers server-level power.
Shared FaaS components such as networking and storage hardware may be divided using our fair division policies. %
The energy costs of keep-alive may also be considered as a separate component in the footprint, and can be modeled based on function eviction rates.
All these expand the energy profiles but require additional power models---current FaasMeter deliberately has a ``minimal'' power model for improving the generalizability and robustness.
We have not achieved full power virtualization---``energy containers'' comprising function or application QoS groups is part of our ongoing work.

\bibliographystyle{acm} %
\bibliography{efaas,faas,mypubs,faas-asplos,energy,carbon}

\begin{thebibliography}{10}

\bibitem{cloudflare-workers}
Cloudflare workers.
\newblock \url{https://blog.cloudflare.com/introducing-cloudflare-workers/}.

\bibitem{gcp_carbon_22}
Google cloud carbon {Footprint}.
\newblock \url{https://cloud.google.com/carbon-footprint}.

\bibitem{azure_sustainability_calc}
Microsoft {Sustainability} {Calculator} helps enterprises analyze the carbon
  emissions of their {IT} infrastructure.
\newblock \url
  {https://azure.microsoft.com/en-us/blog/microsoft-sustainability-calculator-helps-enterprises-analyze-the-carbon-emissions-of-their-it-infrastructure/}.

\bibitem{watttime}
Watttime – {The} {Power} to {Choose} {Clean} {Energy}.
\newblock \url{https://www.watttime.org/}.

\bibitem{openwhisk}
{Apache OpenWhisk: Open Source Serverless Cloud Platform}.
\newblock \url{https://openwhisk.apache.org/}, 2020.

\bibitem{aws-lambda}
{AWS Lambda}.
\newblock \url{https://aws.amazon.com/lambda/}, 2020.

\bibitem{azure-functions}
{Azure Functions}.
\newblock \url{https://azure.microsoft.com/en-us/services/functions/ }, 2020.

\bibitem{google-functions}
{Google Cloud Functions}.
\newblock \url{https://cloud.google.com/functions }, 2020.

\bibitem{aws_carbon_tool}
{Customer} {Carbon} {Footprint} {Tool} {\textbar} {AWS} {News} {Blog}, Mar.
  2022.
\newblock Section: Announcements.

\bibitem{jetson_2022}
Jetson {AGX} {Orin} {Developer} {Kit} {User} {Guide}.
\newblock
  \url{https://developer.nvidia.com/embedded/learn/jetson-agx-orin-devkit-user-guide/index.html},
  Mar. 2022.

\bibitem{acun_carbon_2022}
{\sc Acun, B., Lee, B., Kazhamiaka, F., Maeng, K., Chakkaravarthy, M., Gupta,
  U., Brooks, D., and Wu, C.-J.}
\newblock Carbon {Explorer}: {A} {Holistic} {Approach} for {Designing} {Carbon}
  {Aware} {Datacenters}, May 2022.
\newblock arXiv:2201.10036 [cs, eess].

\bibitem{adzic2017serverless}
{\sc Adzic, G., and Chatley, R.}
\newblock Serverless computing: economic and architectural impact.
\newblock In {\em Proceedings of the 2017 11th Joint Meeting on Foundations of
  Software Engineering\/} (2017), pp.~884--889.

\bibitem{firecracker-nsdi20}
{\sc Agache, A., Brooker, M., Iordache, A., Liguori, A., Neugebauer, R.,
  Piwonka, P., and Popa, D.-M.}
\newblock Firecracker: Lightweight virtualization for serverless applications.
\newblock In {\em 17th {USENIX} Symposium on Networked Systems Design and
  Implementation ({NSDI} 20)\/} (2020), pp.~419--434.

\bibitem{agarwal_redesigning_2021}
{\sc Agarwal, A., Sun, J., Noghabi, S., Iyengar, S., Badam, A., Chandra, R.,
  Seshan, S., and Kalyanaraman, S.}
\newblock Redesigning {Data} {Centers} for {Renewable} {Energy}.
\newblock {\em HotNets\/} (2021), 8.

\bibitem{agmon2014rise}
{\sc Agmon Ben-Yehuda, O., Ben-Yehuda, M., Schuster, A., and Tsafrir, D.}
\newblock The rise of raas: the resource-as-a-service cloud.
\newblock {\em Communications of the ACM 57}, 7 (2014), 76--84.

\bibitem{anand_hotcarbon22}
{\sc Anand, V., Xie, Z., Stolet, M., De~Viti, R., Davidson, T., Karimipour, R.,
  Alzayat, S., and Mace, J.}
\newblock {The Odd One Out: Energy is not like Other Metrics}.
\newblock {\em HotCarbon 2022: 1st Workshop on Sustainable Computer Systems
  Design and Implementation\/} (July 2022).

\bibitem{anderson_treehouse_2022}
{\sc Anderson, T., Belay, A., Chowdhury, M., Cidon, A., and Zhang, I.}
\newblock Treehouse: {A} {Case} {For} {Carbon}-{Aware} {Datacenter} {Software}.
\newblock {\em arXiv:2201.02120 [cs]\/} (Jan. 2022).
\newblock arXiv: 2201.02120.

\bibitem{anthony_carbontracker_2020}
{\sc Anthony, L. F.~W., Kanding, B., and Selvan, R.}
\newblock Carbontracker: {Tracking} and {Predicting} the {Carbon} {Footprint}
  of {Training} {Deep} {Learning} {Models}.
\newblock {\em ICML Workshop on Challenges in Deploying and monitoring Machine
  Learning Systems\/} (July 2020).
\newblock arXiv:2007.03051 [cs, eess, stat].

\bibitem{babakol_calm_2020}
{\sc Babakol, T., Canino, A., Mahmoud, K., Saxena, R., and Liu, Y.~D.}
\newblock Calm energy accounting for multithreaded {Java} applications.
\newblock In {\em Proceedings of the 28th {ACM} {Joint} {Meeting} on {European}
  {Software} {Engineering} {Conference} and {Symposium} on the {Foundations} of
  {Software} {Engineering}\/} (Virtual Event USA, Nov. 2020), ACM,
  pp.~976--988.

\bibitem{bellosa2000benefits}
{\sc Bellosa, F.}
\newblock The benefits of event: driven energy accounting in power-sensitive
  systems.
\newblock In {\em Proceedings of the 9th workshop on ACM SIGOPS European
  workshop: beyond the PC: new challenges for the operating system\/} (2000),
  pp.~37--42.

\bibitem{carreira_cirrus_2019}
{\sc Carreira, J., Fonseca, P., Tumanov, A., Zhang, A., and Katz, R.}
\newblock Cirrus: a {Serverless} {Framework} for {End}-to-end {ML} {Workflows}.
\newblock In {\em Proceedings of the {ACM} {Symposium} on {Cloud} {Computing} -
  {SoCC} '19\/} (Santa Cruz, CA, USA, 2019), ACM Press, pp.~13--24.

\bibitem{carroll2010analysis}
{\sc Carroll, A., and Heiser, G.}
\newblock An analysis of power consumption in a smartphone.
\newblock In {\em 2010 USENIX Annual Technical Conference (USENIX ATC 10)\/}
  (2010).

\bibitem{castro2019rise}
{\sc Castro, P., Ishakian, V., Muthusamy, V., and Slominski, A.}
\newblock The rise of serverless computing.
\newblock {\em Communications of the ACM 62}, 12 (2019), 44--54.

\bibitem{funcx_hpdc_20}
{\sc Chard, R., Babuji, Y., Li, Z., Skluzacek, T., Woodard, A., Blaiszik, B.,
  Foster, I., and Chard, K.}
\newblock Funcx: A federated function serving fabric for science.
\newblock In {\em Proceedings of the 29th International Symposium on
  High-Performance Parallel and Distributed Computing\/} (New York, NY, USA,
  2020), HPDC 20, Association for Computing Machinery, pp.~65--76.

\bibitem{cncf_kepler}
{\sc CNCF}.
\newblock Kepler: {Kubernetes} {Efficient} {Power} {Level} {Exporter}.
\newblock \url{https://sustainable-computing.io/}.

\bibitem{colmant_next_2018}
{\sc Colmant, M., Rouvoy, R., Kurpicz, M., Sobe, A., Felber, P., and
  Seinturier, L.}
\newblock The next 700 {CPU} power models.
\newblock {\em Journal of Systems and Software 144\/} (Oct. 2018), 382--396.

\bibitem{desrochers_validation_2016}
{\sc Desrochers, S., Paradis, C., and Weaver, V.~M.}
\newblock A {Validation} of {DRAM} {RAPL} {Power} {Measurements}.
\newblock In {\em Proceedings of the {Second} {International} {Symposium} on
  {Memory} {Systems}\/} (Alexandria VA USA, Oct. 2016), ACM, pp.~455--470.

\bibitem{do_ptop_2009}
{\sc Do, T., Rawshdeh, S., and Shi, W.}
\newblock {pTop}: {A} {Process}-level {Power} {Profiling} {Tool}.
\newblock {\em HotPower\/} (2009), 5.

\bibitem{dong_rethink_2014}
{\sc Dong, M., Lan, T., and Zhong, L.}
\newblock Rethink energy accounting with cooperative game theory.
\newblock In {\em Proceedings of the 20th annual international conference on
  {Mobile} computing and networking\/} (Maui Hawaii USA, Sept. 2014), ACM,
  pp.~531--542.

\bibitem{du_serverless_2022}
{\sc Du, D., Liu, Q., Jiang, X., Xia, Y., Zang, B., and Chen, H.}
\newblock Serverless computing on heterogeneous computers.
\newblock In {\em Proceedings of the 27th {ACM} {International} {Conference} on
  {Architectural} {Support} for {Programming} {Languages} and {Operating}
  {Systems}\/} (Lausanne Switzerland, Feb. 2022), ACM, pp.~797--813.

\bibitem{fieni_smartwatts_2020}
{\sc Fieni, G., Rouvoy, R., and Seinturier, L.}
\newblock {SmartWatts}: {Self}-{Calibrating} {Software}-{Defined} {Power}
  {Meter} for {Containers}.
\newblock {\em arXiv:2001.02505 [cs]\/} (Jan. 2020).
\newblock arXiv: 2001.02505.

\bibitem{fieni_selfwatts_2021}
{\sc Fieni, G., Rouvoy, R., and Seiturier, L.}
\newblock {SelfWatts}: {On}-the-fly {Selection} of {Performance} {Events} to
  {Optimize} {Software}-defined {Power} {Meters}.
\newblock In {\em 2021 {IEEE}/{ACM} 21st {International} {Symposium} on
  {Cluster}, {Cloud} and {Internet} {Computing} ({CCGrid})\/} (Melbourne,
  Australia, May 2021), IEEE, pp.~324--333.

\bibitem{flinn_energy-aware_1999}
{\sc Flinn, J., and Satyanarayanan, M.}
\newblock Energy-aware adaptation for mobile applications.
\newblock {\em ACM SIGOPS Operating Systems Review\/} (1999), 16.

\bibitem{flinn_powerscope_1999}
{\sc Flinn, J., and Satyanarayanan, M.}
\newblock {PowerScope}: a tool for profiling the energy usage of mobile
  applications.
\newblock In {\em Proceedings {WMCSA}'99. {Second} {IEEE} {Workshop} on
  {Mobile} {Computing} {Systems} and {Applications}\/} (Feb. 1999), pp.~2--10.

\bibitem{fonseca_quanto_2008}
{\sc Fonseca, R., Dutta, P., Levis, P., and Stoica, I.}
\newblock Quanto: {Tracking} {Energy} in {Networked} {Embedded} {Systems}.
\newblock {\em OSDI\/} (2008), 16.

\bibitem{fouladi_laptop_2019}
{\sc Fouladi, S., Romero, F., Iter, D., Li, Q., and Chatterjee, S.}
\newblock From {Laptop} to {Lambda}: {Outsourcing} {Everyday} {Jobs} to
  {Thousands} of {Transient} {Functional} {Containers}.
\newblock {\em USENIX Annual Technical Conference\/} (2019), 15.

\bibitem{fouladi2017encoding}
{\sc Fouladi, S., Wahby, R.~S., Shacklett, B., Balasubramaniam, K.~V., Zeng,
  W., Bhalerao, R., Sivaraman, A., Porter, G., and Winstein, K.}
\newblock Encoding, fast and slow: Low-latency video processing using thousands
  of tiny threads.
\newblock In {\em 14th USENIX Symposium on Networked Systems Design and
  Implementation (NSDI 17)\/} (2017), pp.~363--376.

\bibitem{il76-hpdc23}
{\sc Fuerst, A., Rehman, A., and Sharma, P.}
\newblock Il\'{u}vatar: A fast control plane for serverless computing.
\newblock In {\em Proceedings of the 32nd {International} {Symposium} on
  {High}-{Performance} {Parallel} and {Distributed} {Computing}\/} (June 2023),
  {HPDC} '23, Association for Computing Machinery.

\bibitem{fuerst2023iluvatar}
{\sc Fuerst, A., Rehman, A., and Sharma, P.}
\newblock Il{\'u}vatar: A fast control plane for serverless computing.

\bibitem{faascache-asplos21}
{\sc Fuerst, A., and Sharma, P.}
\newblock Faascache: Keeping serverless computing alive with greedy-dual
  caching.
\newblock In {\em Proceedings of the 26th ACM International Conference on
  Architectural Support for Programming Languages and Operating Systems\/} (New
  York, NY, USA, 2021), ASPLOS 2021, Association for Computing Machinery,
  pp.~386--400.

\bibitem{ge_powerpack_2010}
{\sc Ge, R., Feng, X., Song, S., Chang, H.-C., Li, D., and Cameron, K.~W.}
\newblock {PowerPack}: {Energy} {Profiling} and {Analysis} of
  {High}-{Performance} {Systems} and {Applications}.
\newblock {\em IEEE Transactions on Parallel and Distributed Systems 21}, 5
  (May 2010), 658--671.
\newblock Conference Name: IEEE Transactions on Parallel and Distributed
  Systems.

\bibitem{ghanei_os-based_2016}
{\sc Ghanei, F., Tipnis, P., Marcus, K., Dantu, K., Ko, S., and Ziarek, L.}
\newblock {OS}-based {Resource} {Accounting} for {Asynchronous} {Resource}
  {Use} in {Mobile} {Systems}.
\newblock In {\em Proceedings of the 2016 {International} {Symposium} on {Low}
  {Power} {Electronics} and {Design}\/} (San Francisco Airport CA USA, Aug.
  2016), ACM, pp.~296--301.

\bibitem{ghanei_os-based_2019}
{\sc Ghanei, F., Tipnis, P., Marcus, K., Dantu, K., Ko, S.~Y., and Ziarek, L.}
\newblock {OS}-{Based} {Energy} {Accounting} for {Asynchronous} {Resources} in
  {IoT} {Devices}.
\newblock {\em IEEE Internet of Things Journal 6}, 3 (June 2019), 5841--5852.
\newblock Conference Name: IEEE Internet of Things Journal.

\bibitem{guo_power_2018}
{\sc Guo, L., Xu, T., Xu, M., Liu, X., and Lin, F.~X.}
\newblock Power sandbox: power awareness redefined.
\newblock In {\em Proceedings of the {Thirteenth} {EuroSys} {Conference}\/}
  (Porto Portugal, Apr. 2018), ACM, pp.~1--15.

\bibitem{henderson_towards_2020}
{\sc Henderson, P., Hu, J., Romoff, J., Brunskill, E., Jurafsky, D., and
  Pineau, J.}
\newblock Towards the {Systematic} {Reporting} of the {Energy} and {Carbon}
  {Footprints} of {Machine} {Learning}.
\newblock {\em Journal of Machine Learning Research 21\/} (2020), 1--43.

\bibitem{scaph}
{\sc Hubblo}.
\newblock Scaphandre.
\newblock \url{https://github.com/hubblo-org/scaphandre}, July 2023.

\bibitem{he_energat_2023}
{\sc Hè, H., Friedman, M., and Rekatsinas, T.}
\newblock {EnergAt}: {Fine}-{Grained} {Energy} {Attribution} for
  {Multi}-{Tenancy}.
\newblock {\em HotCarbon\/} (2023).

\bibitem{islam2016new}
{\sc Islam, M.~A., and Ren, S.}
\newblock A new perspective on energy accounting in $\{$Multi-Tenant$\}$ data
  centers.
\newblock In {\em USENIX Workshop on Cool Topics on Sustainable Data Centers
  (CoolDC 16)\/} (2016).

\bibitem{jay_experimental_2023}
{\sc Jay, M., Ostapenco, V., Lefevre, L., Trystram, D., Orgerie, A.-C., and
  Fichel, B.}
\newblock An experimental comparison of software-based power meters: focus on
  {CPU} and {GPU}.
\newblock In {\em 2023 {IEEE}/{ACM} 23rd {International} {Symposium} on
  {Cluster}, {Cloud} and {Internet} {Computing} ({CCGrid})\/} (May 2023),
  pp.~106--118.

\bibitem{jiang_virtual_2017}
{\sc Jiang, W., Liu, F., Tang, G., Wu, K., and Jin, H.}
\newblock Virtual {Machine} {Power} {Accounting} with {Shapley} {Value}.
\newblock In {\em 2017 {IEEE} 37th {International} {Conference} on
  {Distributed} {Computing} {Systems} ({ICDCS})\/} (Atlanta, GA, USA, June
  2017), IEEE, pp.~1683--1693.

\bibitem{jiang_non-it_2018}
{\sc Jiang, W., Ren, S., Liu, F., and Jin, H.}
\newblock Non-{IT} {Energy} {Accounting} in {Virtualized} {Datacenter}.
\newblock In {\em 2018 {IEEE} 38th {International} {Conference} on
  {Distributed} {Computing} {Systems} ({ICDCS})\/} (Vienna, July 2018), IEEE,
  pp.~300--310.

\bibitem{khan_rapl_2018}
{\sc Khan, K.~N., Hirki, M., Niemi, T., Nurminen, J.~K., and Ou, Z.}
\newblock {RAPL} in {Action}: {Experiences} in {Using} {RAPL} for {Power}
  {Measurements}.
\newblock {\em ACM Transactions on Modeling and Performance Evaluation of
  Computing Systems 3}, 2 (June 2018), 1--26.

\bibitem{khan_energy_2015}
{\sc Khan, K.~N., Nyback, F., Ou, Z., Nurminen, J.~K., Niemi, T., Eulisse, G.,
  Elmer, P., and Abdurachmanov, D.}
\newblock Energy {Profiling} {Using} {IgProf}.
\newblock In {\em 2015 15th {IEEE}/{ACM} {International} {Symposium} on
  {Cluster}, {Cloud} and {Grid} {Computing}\/} (Shenzhen, China, May 2015),
  IEEE, pp.~1115--1118.

\bibitem{kim_functionbench_2019}
{\sc Kim, J., and Lee, K.}
\newblock {FunctionBench}: {A} {Suite} of {Workloads} for {Serverless} {Cloud}
  {Function} {Service}.
\newblock In {\em 2019 {IEEE} 12th {International} {Conference} on {Cloud}
  {Computing} ({CLOUD})\/} (July 2019), pp.~502--504.
\newblock ISSN: 2159-6182.

\bibitem{lee_cloudsocket_2018}
{\sc Lee, S., Kim, H., Park, S., Kim, S., Choe, H., and Yoon, S.}
\newblock {CloudSocket}: {Fine}-{Grained} {Power} {Sensing} {System} for
  {Datacenters}.
\newblock {\em IEEE Access 6\/} (2018), 49601--49610.
\newblock Conference Name: IEEE Access.

\bibitem{mahgoub_wisefuse_2022}
{\sc Mahgoub, A., Yi, E.~B., Shankar, K., Minocha, E., Elnikety, S., Bagchi,
  S., and Chaterji, S.}
\newblock {WISEFUSE}: {Workload} {Characterization} and {DAG} {Transformation}
  for {Serverless} {Workflows}.
\newblock {\em Proceedings of the ACM on Measurement and Analysis of Computing
  Systems 6}, 2 (May 2022), 1--28.

\bibitem{maji2022carboncast}
{\sc Maji, D., Shenoy, P., and Sitaraman, R.~K.}
\newblock Carboncast: multi-day forecasting of grid carbon intensity.
\newblock In {\em Proceedings of the 9th ACM International Conference on
  Systems for Energy-Efficient Buildings, Cities, and Transportation\/} (2022),
  pp.~198--207.

\bibitem{maschi_serverless_2023}
{\sc Maschi, F., Korolija, D., and Alonso, G.}
\newblock Serverless {FPGA}: {Work}-{In}-{Progress}.
\newblock In {\em Proceedings of the 1st {Workshop} on {SErverless} {Systems},
  {Applications} and {MEthodologies}\/} (Rome Italy, May 2023), ACM, pp.~1--4.

\bibitem{meinhold1983understanding}
{\sc Meinhold, R.~J., and Singpurwalla, N.~D.}
\newblock {Understanding the Kalman filter}.
\newblock {\em The American Statistician 37}, 2 (1983), 123--127.

\bibitem{mukhanov_alea_2017}
{\sc Mukhanov, L., Petoumenos, P., Wang, Z., Parasyris, N., Nikolopoulos,
  D.~S., De~Supinski, B.~R., and Leather, H.}
\newblock {ALEA}: {A} {Fine}-{Grained} {Energy} {Profiling} {Tool}.
\newblock {\em ACM Transactions on Architecture and Code Optimization 14}, 1
  (Apr. 2017), 1--25.

\bibitem{narayan2013power}
{\sc Narayan, A., and Rao, S.}
\newblock Power-aware cloud metering.
\newblock {\em IEEE Transactions on Services Computing 7}, 3 (2013), 440--451.

\bibitem{noureddine2013review}
{\sc Noureddine, A., Rouvoy, R., and Seinturier, L.}
\newblock A review of energy measurement approaches.
\newblock {\em ACM SIGOPS Operating Systems Review 47}, 3 (2013), 42--49.

\bibitem{oliner2013carat}
{\sc Oliner, A.~J., Iyer, A.~P., Stoica, I., Lagerspetz, E., and Tarkoma, S.}
\newblock Carat: Collaborative energy diagnosis for mobile devices.
\newblock In {\em Proceedings of the 11th ACM conference on embedded networked
  sensor systems\/} (2013), pp.~1--14.

\bibitem{ournani2020taming}
{\sc Ournani, Z., Belgaid, M.~C., Rouvoy, R., Rust, P., Penhoat, J., and
  Seinturier, L.}
\newblock Taming energy consumption variations in systems benchmarking.
\newblock In {\em Proceedings of the ACM/SPEC International Conference on
  Performance Engineering\/} (2020), pp.~36--47.

\bibitem{pathak2012energy}
{\sc Pathak, A., Hu, Y.~C., and Zhang, M.}
\newblock Where is the energy spent inside my app? fine grained energy
  accounting on smartphones with eprof.
\newblock In {\em Proceedings of the 7th ACM european conference on Computer
  Systems\/} (2012), pp.~29--42.

\bibitem{Pearson2019}
{\sc Pearson, K.~A., Griffith, C.~A., Zellem, R.~T., Koskinen, T.~T., and
  Roudier, G.~M.}
\newblock Ground-based spectroscopy of the exoplanet xo-2b using a systematic
  wavelength calibration.
\newblock {\em The Astronomical Journal 157}, 1 (2019), 21.

\bibitem{scikit-learn}
{\sc Pedregosa, F., Varoquaux, G., Gramfort, A., Michel, V., Thirion, B.,
  Grisel, O., Blondel, M., Prettenhofer, P., Weiss, R., Dubourg, V.,
  Vanderplas, J., Passos, A., Cournapeau, D., Brucher, M., Perrot, M., and
  Duchesnay, E.}
\newblock Scikit-learn: Machine learning in {P}ython.
\newblock {\em Journal of Machine Learning Research 12\/} (2011), 2825--2830.

\bibitem{rastegar_enex_2023}
{\sc Rastegar, S.~H., Shafiei, H., and Khonsari, A.}
\newblock {EneX}: {An} {Energy}-{Aware} {Execution} {Scheduler} for
  {Serverless} {Computing}.
\newblock {\em IEEE Transactions on Industrial Informatics\/} (2023), 1--13.

\bibitem{schirmer_night_2023}
{\sc Schirmer, T., Japke, N., Greten, S., Pfandzelter, T., and Bermbach, D.}
\newblock The {Night} {Shift}: {Understanding} {Performance} {Variability} of
  {Cloud} {Serverless} {Platforms}.
\newblock In {\em Proceedings of the 1st {Workshop} on {SErverless} {Systems},
  {Applications} and {MEthodologies}\/} (Rome Italy, May 2023), ACM,
  pp.~27--33.

\bibitem{serverless-cacm-21}
{\sc Schleier-Smith, J., Sreekanti, V., Khandelwal, A., Carreira, J.,
  Yadwadkar, N.~J., Popa, R.~A., Gonzalez, J.~E., Stoica, I., and Patterson,
  D.~A.}
\newblock What serverless computing is and should become: The next phase of
  cloud computing.
\newblock {\em Commun. ACM 64}, 5 (Apr. 2021), 76–84.

\bibitem{schmitt_online_2019}
{\sc Schmitt, N., Iffländer, L., Bauer, A., and Kounev, S.}
\newblock Online {Power} {Consumption} {Estimation} for {Functions} in {Cloud}
  {Applications}.
\newblock In {\em 2019 {IEEE} {International} {Conference} on {Autonomic}
  {Computing} ({ICAC})\/} (June 2019), pp.~63--72.
\newblock ISSN: 2474-0756.

\bibitem{sklearnsvmsvr}
{\sc Scikit-learn}.
\newblock Support vector regression.
\newblock
  \url{https://scikit-learn/stable/modules/generated/sklearn.svm.SVR.html}.

\bibitem{shahrad_serverless_2020}
{\sc Shahrad, M., Fonseca, R., Goiri, I., Chaudhry, G., Batum, P., Cooke, J.,
  Laureano, E., Tresness, C., Russinovich, M., and Bianchini, R.}
\newblock {Serverless in the Wild: Characterizing and Optimizing the Serverless
  Workload at a Large Cloud Provider}.
\newblock In {\em 2020 USENIX annual technical conference (USENIX ATC 20)\/}
  (2020), pp.~205--218.

\bibitem{sharma_hotcarbon22}
{\sc Sharma, P.}
\newblock Challenges and opportunities in sustainable serverless computing.
\newblock {\em HotCarbon 2022: 1st Workshop on Sustainable Computer Systems
  Design and Implementation\/} (July 2022).

\bibitem{mghpcc-ic}
{\sc Sharma, P., Pegus, P.~I., Irwin, D., Shenoy, P., Goodhue, J., and Culbert,
  J.}
\newblock Design and operational analysis of a green data center.
\newblock {\em IEEE Internet Computing 21}, 4 (2017), 16--24.

\bibitem{shen_power_2013}
{\sc Shen, K., Shriraman, A., Dwarkadas, S., Zhang, X., and Chen, Z.}
\newblock Power containers: an {OS} facility for fine-grained power and energy
  management on multicore servers.
\newblock {\em ASPLOS\/} (2013), 12.

\bibitem{Siddik_2021}
{\sc Siddik, M. A.~B., Shehabi, A., and Marston, L.}
\newblock The environmental footprint of data centers in the united states.
\newblock {\em Environmental Research Letters 16}, 6 (may 2021), 064017.

\bibitem{souza2023ecovisor}
{\sc Souza, A., Bashir, N., Murillo, J., Hanafy, W., Liang, Q., Irwin, D., and
  Shenoy, P.}
\newblock Ecovisor: A virtual energy system for carbon-efficient applications.
\newblock In {\em Proceedings of the 28th ACM International Conference on
  Architectural Support for Programming Languages and Operating Systems, Volume
  2\/} (2023), pp.~252--265.

\bibitem{tzenetopoulos_dvfaas_2023}
{\sc Tzenetopoulos, A., Masouros, D., Soudris, D., and Xydis, S.}
\newblock {DVFaaS}: {Leveraging} {DVFS} for {FaaS} workflows.
\newblock {\em IEEE Computer Architecture Letters\/} (2023), 1--4.

\bibitem{ustiugov_analyzing_2021}
{\sc Ustiugov, D., Amariucai, T., and Grot, B.}
\newblock Analyzing {Tail} {Latency} in {Serverless} {Clouds} with {STeLLAR}.
\newblock In {\em 2021 {IEEE} {International} {Symposium} on {Workload}
  {Characterization} ({IISWC})\/} (Storrs, CT, USA, Nov. 2021), IEEE,
  pp.~51--62.

\bibitem{vergara_sharing_2015}
{\sc Vergara, E.~J., Nadjm-Tehrani, S., and Asplund, M.}
\newblock Sharing the {Cost} of {Lunch}: {Energy} {Apportionment} {Policies}.
\newblock In {\em Proceedings of the 11th {ACM} {Symposium} on {QoS} and
  {Security} for {Wireless} and {Mobile} {Networks}\/} (Cancun Mexico, Nov.
  2015), ACM, pp.~91--97.

\bibitem{wagner_energy_2023}
{\sc Wagner, L., Mayer, M., Marino, A., Soldani~Nezhad, A., Zwaan, H., and
  Malavolta, I.}
\newblock On the {Energy} {Consumption} and {Performance} of {WebAssembly}
  {Binaries} across {Programming} {Languages} and {Runtimes} in {IoT}.
\newblock In {\em Proceedings of the 27th {International} {Conference} on
  {Evaluation} and {Assessment} in {Software} {Engineering}\/} (Oulu Finland,
  June 2023), ACM, pp.~72--82.

\bibitem{wang2021lass}
{\sc Wang, B., Ali-Eldin, A., and Shenoy, P.}
\newblock Lass: Running latency sensitive serverless computations at the edge.
\newblock In {\em Proceedings of the 30th international symposium on
  high-performance parallel and distributed computing\/} (2021), pp.~239--251.

\bibitem{wang_real-time_2020}
{\sc Wang, R., Van~Le, D., Tan, R., Wong, Y.-W., and Wen, Y.}
\newblock Real-{Time} {Cooling} {Power} {Attribution} for {Co}-{Located} {Data}
  {Center} {Rooms} with {Distinct} {Temperatures}.
\newblock In {\em Proceedings of the 7th {ACM} {International} {Conference} on
  {Systems} for {Energy}-{Efficient} {Buildings}, {Cities}, and
  {Transportation}\/} (Virtual Event Japan, Nov. 2020), ACM, pp.~190--199.

\bibitem{welch1995introduction}
{\sc Welch, G., Bishop, G., et~al.}
\newblock {\em An introduction to the Kalman filter}.
\newblock Chapel Hill, NC, USA, 1995.

\bibitem{winter2002shapley}
{\sc Winter, E.}
\newblock The shapley value.
\newblock {\em Handbook of game theory with economic applications 3\/} (2002),
  2025--2054.

\bibitem{xu2021lambda}
{\sc Xu, F., Qin, Y., Chen, L., Zhou, Z., and Liu, F.}
\newblock $\lambda$-dnn : Achieving predictable distributed dnn training with
  serverless architectures.
\newblock {\em IEEE Transactions on Computers\/} (2021).

\bibitem{yoon_appscope_2012}
{\sc Yoon, C., Kim, D., Jung, W., Kang, C., and Cha, H.}
\newblock {AppScope}: {Application} {Energy} {Metering} {Framework} for
  {Android} {Smartphones} using {Kernel} {Activity} {Monitoring}.
\newblock {\em USENIX ATC\/} (2012), 14.

\bibitem{yu_faaswap_2023}
{\sc Yu, M., Wang, A., Chen, D., Yu, H., Luo, X., Li, Z., Wang, W., Chen, R.,
  Nie, D., and Yang, H.}
\newblock {FaaSwap}: {SLO}-{Aware}, {GPU}-{Efficient} {Serverless} {Inference}
  via {Model} {Swapping}, June 2023.
\newblock arXiv:2306.03622 [cs].

\bibitem{zeng_ecosystem_2002}
{\sc Zeng, H., Ellis, C.~S., Lebeck, A.~R., and Vahdat, A.}
\newblock {ECOSystem}: {Managing} {Energy} as a {First} {Class} {Operating}
  {System} {Resource}.
\newblock {\em ASPLOS\/} (2002), 10.

\bibitem{zeng_currentcy_2003}
{\sc Zeng, H., Ellis, C.~S., Lebeck, A.~R., and Vahdat, A.}
\newblock Currentcy: {A} {Unifying} {Abstraction} for {Expressing} {Energy}
  {Management} {Policies}.
\newblock {\em USENIX ATC\/} (2003), 14.

\bibitem{zhang_quantitative_2015}
{\sc Zhang, H., and Hoffmann, H.}
\newblock A {Quantitative} {Evaluation} of the {RAPL} {Power} {Control}
  {System}.
\newblock {\em Feedback computing\/} (2015), 6.

\bibitem{zhang_maximizing_2016}
{\sc Zhang, H., and Hoffmann, H.}
\newblock Maximizing {Performance} {Under} a {Power} {Cap}: {A} {Comparison} of
  {Hardware}, {Software}, and {Hybrid} {Techniques}.
\newblock {\em ASPLOS\/} (2016), 15.

\bibitem{zhang_estimating_2020}
{\sc Zhang, X., Shen, Z., Xia, B., Liu, Z., and Li, Y.}
\newblock Estimating {Power} {Consumption} of {Containers} and {Virtual}
  {Machines} in {Data} {Centers}.
\newblock In {\em 2020 {IEEE} {International} {Conference} on {Cluster}
  {Computing} ({CLUSTER})\/} (Sept. 2020), pp.~288--293.
\newblock ISSN: 2168-9253.

\end{thebibliography}
\newpage

\end{document}